\documentclass[prd,twocolumn,preprintnumbers,superscriptaddress,nofootinbib]{revtex4}
\usepackage[a4paper, hdivide={1.91cm,,1.165cm}, vdivide={1.83cm,,3.3cm}]{geometry}

\usepackage{amsmath,amssymb}
\usepackage{graphicx}
\usepackage{dcolumn}
\usepackage[hyperfootnotes=false]{hyperref}
\usepackage{xspace}
\usepackage{color}

\newcommand{\eq}[1]{Eq.~\eqref{#1}}
 
\newcommand{\MeV}{\,\text{MeV}}
\newcommand{\GeV}{\,\text{GeV}}
\newcommand{\TeV}{\,\text{TeV}}
\newcommand{\eV}{\,\text{eV}}
\newcommand{\meV}{\,\text{meV}}

\newcommand{\diag}{\text{diag}}
\newcommand{\BKs}{$B\to K^*\mu^+\mu^-$\ }

\def \epsilon {\varepsilon} 
\newcommand{\matrixx}[1]{\begin{pmatrix} #1 \end{pmatrix}} 
\newcommand{\hc}{\ensuremath{\text{h.c.}}}

\allowdisplaybreaks 

\begin{document}
\preprint{\vbox{\hbox{CERN-PH-TH-2015-046}\hbox{ULB-TH/15-03}}}

\title{Addressing the LHC flavour anomalies with horizontal gauge symmetries}

\author{Andreas Crivellin}
\affiliation{CERN Theory Division, CH--1211 Geneva 23, Switzerland}%
\author{Giancarlo D'Ambrosio}
\affiliation{CERN Theory Division, CH--1211 Geneva 23, Switzerland}%
\affiliation{INFN-Sezione di Napoli, Via Cintia, 80126 Napoli, Italy}
\author{Julian Heeck}
\affiliation{Service de Physique Th\'eorique, Universit\'e Libre de Bruxelles, Boulevard du Triomphe, CP225, 1050 Brussels, Belgium}

\begin{abstract}
We study the impact of an additional $U(1)'$ gauge symmetry with flavour-dependent charges for quarks and leptons on the LHC flavour anomalies observed in $B\to K^* \mu^+\mu^-$, $R(K)=B\to K \mu^+\mu^-/B\to K e^+e^-$, and $h\to\mu\tau$. In its minimal version with two scalar doublets, the resulting model naturally explains the deviations from the Standard Model observed in $B\to K^* \mu^+\mu^-$ and $R(K)$. The CMS access in $h\to\mu\tau$ can be explained by introducing a third scalar doublet, which gives rise to a prediction for $\tau\to 3\mu$. We investigate constraints from flavour observables and direct LHC searches for $pp\to Z' \to \mu^+\mu^-$. 
Our model successfully generates the measured fermion-mixing matrices and does not require vector-like fermions, unlike previous attempts to explain these anomalies.
\end{abstract}


\maketitle

\section{Introduction}
\label{intro}

The discovery of a scalar particle at the LHC~\cite{Aad:2012tfa,Chatrchyan:2012ufa} with properties close to its theoretical prediction within the Standard Model (SM) marks its completion as a description of particle physics. While direct searches for physics beyond the SM were negative at the first LHC run, there are some interesting indirect hints for new physics effects in the flavour sector, namely in the decays of $B$ mesons -- $B\to K^* \mu^+\mu^-$ and $R(K)=B\to K \mu^+\mu^-/B\to K e^+e^-$ -- and in the decay of the Brout--Englert--Higgs boson $h\to\mu\tau$.

Specifically, the deviations from the SM found by LHCb~\cite{Aaij:2013qta} in the decay $B\to K^* \mu^+\mu^-$ arise mainly in an angular observable called $P_5^\prime$~\cite{Descotes-Genon:2013vna}, with a significance of $2$--$3\,\sigma$ depending on the assumptions for the hadronic uncertainties~\cite{Descotes-Genon:2014uoa,Altmannshofer:2014rta,Jager:2014rwa}. This effect can be explained in a model-independent effective-field-theory approach by a fairly large contribution of the operator $C_9^{\mu\mu}(\overline{s}\gamma_\alpha P_L b)(\overline{\mu}\gamma^\alpha \mu)$~\cite{Descotes-Genon:2013wba,Altmannshofer:2013foa,Horgan:2013pva}.\footnote{According to Ref.~\cite{Lyon:2014hpa}, also underestimated charm effects could explain the deviations from the SM.}
LHCb further observed lepton-non-universality in the $B$-meson decays~\cite{Aaij:2014ora}
\begin{equation}
	R(K)=\frac{B\to K \mu^+\mu^-}{B\to K e^+e^-}=0.745^{+0.090}_{-0.074}\pm 0.036\,,
\end{equation}
which deviates from the SM prediction $R_K^{\rm SM}=1.0003 \pm 0.0001$~\cite{Bobeth:2007dw} by $2.6\,\sigma$. 
A possible explanation comes again in the form of a non-zero new-physics contribution to $C_9^{\mu\mu}$ -- of the same magnitude as the one required by $B\to K^* \mu^+\mu^-$ \cite{Hurth:2014vma,Altmannshofer:2014rta} -- as long as the analogous contribution to the corresponding operator with electrons, $C_9^{ee}$, is small~\cite{Alonso:2014csa,Hiller:2014yaa,Ghosh:2014awa,Bhattacharya:2014wla}. 

CMS recently presented the results of a search for a lepton-flavour violating (LFV) decay mode of the $125\GeV$ scalar $h\to\mu\tau$, with a preferred value~\cite{Khachatryan:2015kon}
\begin{equation}
	{\rm Br} [h\to\mu\tau] = \left( 0.84_{-0.37}^{+0.39} \right)\% \,,
	\label{h0taumuExp}
\end{equation}
updating an earlier preliminary result~\cite{CMS:2014hha}.
Since this decay is forbidden in the SM it corresponds to a $2.4 \,\sigma$ deviation. This is particularly exciting because it hints at LFV in the charged-lepton sector, whereas we have so far only observed LFV in the neutrino sector through oscillations. Seeing as the simplest SM extensions that can account for neutrino masses and mixing would not lead to observable $h\to\mu\tau$ rates, the confirmation of this decay would have a huge impact on our understanding of lepton flavour. In particular, it would imply potentially measurable rates for LFV processes such as $\tau\to 3\mu$ or $\tau \to \mu\gamma$~\cite{Harnik:2012pb,Blankenburg:2012ex,Davidson:2012ds,Arhrib:2012ax,Arhrib:2012mg,Falkowski:2013jya,Celis:2014roa,Kopp:2014rva,deLima:2015pqa}.
Models aiming to accommodate or explain \eq{h0taumuExp} rely on an extended scalar sector~\cite{Campos:2014zaa,Sierra:2014nqa,Heeck:2014qea,Crivellin:2015mga,Dorsner:2015mja,Omura:2015nja} or non-renormalizable effective operators~\cite{Dery:2014kxa,Lee:2014rba,Dorsner:2015mja}.

An explanation for $h\to\mu\tau$ typically requires additional scalars, while an explanation for $B\to K^* \mu^+\mu^-$ requires additional $Z'$ vector bosons (or leptoquarks \cite{Hiller:2014yaa,Gripaios:2014tna,Sahoo:2015wya}) to generate the current--current interaction $(\overline{s}\gamma_\alpha P_L b)(\overline{\mu}\gamma^\alpha \mu)$~\cite{Gauld:2013qba,Buras:2013qja,Gauld:2013qja,Buras:2013dea}. If the $Z'$ couples non-universally to leptons, it can also account for $R(K)=B\to K \mu^+\mu^-/B\to K e^+e^-$~\cite{Altmannshofer:2014cfa}. 
In Ref.~\cite{Crivellin:2015mga} we presented a model that can resolve all three anomalies, by combining the model of Ref.~\cite{Altmannshofer:2014cfa} (with gauged $U(1)_{L_\mu-L_\tau}$ and effective $Z'$ couplings to quarks generated by heavy vector-quarks) with the one of Ref.~\cite{Heeck:2014qea} (with gauged $U(1)_{L_\mu-L_\tau}$broken by a second scalar doublet).
The combined resolution of the three flavour anomalies gave in particular rise to a prediction for the rate of $\tau\to 3\mu$.

In this article we want to study the possibility that the same effect employed in Refs.~\cite{Heeck:2014qea,Crivellin:2015mga} in the lepton sector might also be responsible for flavour violation in the quark sector. Therefore, also the quarks must be charged under the new $U(1)$ gauge group, leading to a model with flavour dependent $B$ and $L$ charges~\cite{Langacker:2000ju,Barger:2003hg,Cheung:2006tm,Barger:2009eq,Barger:2009qs}. In this way, the introduction of vector-like quarks -- which are somewhat ``artificially'' charged under $L_\mu-L_\tau$ -- can be avoided. Furthermore, the model can explain the smallness of the $V_{ub}$ and $V_{cb}$ elements of the Cabibbo--Kobayashi--Maskawa (CKM) mixing matrix. Since the flavour diagonal couplings to quarks are not arbitrary anymore, interesting correlations with LHC searches arise.

The outline of the article is as follows: In the next section, we will consider the possible charge assignments and the symmetry breaking in the quark sector. Sec.~\ref{flavour} is devoted to a phenomenological analysis of the effects in quark flavour physics and direct LHC searches. Sec.~\ref{sec:3HDM} extends the symmetry breaking to the lepton sector, allowing for a simultaneous explanation of $h\to\mu\tau$. Finally we conclude in Sec.~\ref{sec:conclusion}. In App.~\ref{sec:appendix} we briefly discuss related horizontal gauge symmetries.

\section{Minimal model with two scalar doublets}
\label{sec:2HDM}

Here we study the minimal model with flavour-dependent $U(1)'$ charges which can give rise to the desired effects in $B\to K^* \mu^+\mu^-$ and $R(K)$. To generate the masses and CKM angles, at least two scalar doublets are necessary.

\subsection{Charge assignment}

Concerning leptons, we are drawn to the abelian symmetry $U(1)_{L_\mu-L_\tau}$. It is an anomaly-free global symmetry within the SM~\cite{He:1990pn,Foot:1990mn,He:1991qd} and also a good zeroth-order approximation for neutrino mixing with a quasi-degenerate mass spectrum, predicting a maximal atmospheric but a vanishing reactor neutrino mixing angle~\cite{Binetruy:1996cs,Bell:2000vh,Choubey:2004hn}. Furthermore, since the $Z'$ boson does not couple to electrons, i.e.~$C_9^{ee}=0$, one naturally obtains an effect of the appropriate size in $R(K)$ once $C_9^{\mu\mu}$ acquires its preferred value from $B\to K^*\mu\mu$. Therefore, we choose the following assignment for the charges $Q'$ of the new $U(1)'$ gauge group for the lepton generations:
\begin{equation}
	Q'(L)= (0,\,1,\,-1 )\,.
\end{equation}
Breaking $L_\mu-L_\tau$ is mandatory for a realistic neutrino sector, and such a breaking can also induce charged LFV processes~\cite{Heeck:2011wj}, such as $h\to \mu \tau$ and $\tau\to3\mu$~\cite{Dutta:1994dx,Heeck:2014qea,Crivellin:2015mga}. However, we postpone the discussion of the symmetry breaking in the charged lepton sector to Sec.~\ref{sec:3HDM}. 

Concerning the quark sector, the first two generations should have the same charges in order to avoid very large effects in $K$--$\overline{K}$ or $D$--$\overline{D}$ mixing, generated otherwise unavoidably due to the breaking of the symmetry necessary to generate the Cabibbo angle of the CKM matrix.
Furthermore, the first two generations mix much more strongly among themselves than with the third generation, so the latter seems to be somewhat special. If we require in addition the absence of anomalies, we arrive at the following charge assignment for baryons:
\begin{equation}
	Q'(B)= (-a,\,-a,\,2a )\,.
\end{equation}
We will later study the phenomenological implications of different values of $a$. To reiterate, the $U(1)'$ gauge symmetry we consider is generated by\footnote{Gauge symmetries that couple to $B_1+B_2 - 2 B_3$ have also been discussed in Ref.~\cite{Liu:2011dh} with a focus on effects in the up-quark sector and coupled in the lepton sector not to $L_\mu-L_\tau$, but to the lepton charge $L_e+L_\mu-2 L_\tau$ (which is not a good symmetry in the lepton sector).}
\begin{align}
Q' = (L_\mu - L_\tau) - a (B_1 + B_2 - 2 B_3) \,,
\label{eq:generator}
\end{align}
so the charges are
\begin{align}
Q' &= -\frac{a}{3} \text{ for } u,d,c,s\,, \\
Q' &= \frac{2 a}{3} \text{ for } t,b \,,\\ 
Q' &=  0 \text{ for } e,\nu_e \,, \\
Q' &=  1 \text{ for }\mu,\nu_\mu \,,\text{ and } \\
Q' &= -1 \text{ for }\tau,\nu_\tau\,.
\end{align}
Note that the relative coupling strength to quarks and leptons (parametrized by $a$) is a free parameter because $L_\mu-L_\tau$ and $B_1+B_2-2 B_3$ are independently anomaly free. $a\in\mathbb{Q}$ is nevertheless necessary to avoid massless Goldstone bosons (see below).
Although not required by anomaly cancellation, we also introduce three right-handed neutrinos $\nu_{R}$ with charges $Q'(\nu_R) = (0,\,1,\,-1)$ to employ a seesaw mechanism for neutrino masses.
Other horizontal gauge symmetries can be considered following the above reasoning, but prove less useful in explaining the LHCb anomalies (see App.~\ref{sec:appendix}).

\subsection{Scalar sector}

Two $SU(2)_L$ scalar doublets $\Psi_1$ and $\Psi_2$ are introduced to generate viable fermion masses and quark mixing incorporated in the CKM matrix. They carry the $U(1)'$ charges
\begin{align}
Q'(\Psi_1)=- a\;&& {\rm and}\;&& Q'(\Psi_2)=0\,,
\end{align}
respectively. $\Psi_2$, being uncharged under the $U(1)'$ symmetry, will generate the diagonal entries of the fermion mass matrices as well as the Cabibbo angle. The vacuum expectation value (VEV) $\langle \Psi_1\rangle$ will generate the mixing between the third and first two quark generations necessary for a viable CKM matrix. 

In addition (at least) two SM singlet scalars $\Phi_1$ and $\Phi_2$ with charges
\begin{align}
Q'(\Phi_1)= 1\,, \qquad Q'(\Phi_2)=-a
\label{eq:singlets}
\end{align}
have to be introduced to break the $U(1)'$ gauge symmetry above the electroweak scale and generate the coupling $\Psi_1^\dagger \Psi_2$ necessary for the mixing of the doublets. The VEV $\langle \Phi_1\rangle$ will break the $L_\mu-L_\tau$ symmetry in the right-handed neutrino mass matrix relevant for the seesaw mechanism, which leads to a valid neutrino mixing matrix~\cite{Araki:2012ip,Heeck:2014qea}. $\Phi_2 $ generates the term $\langle \Phi_2\rangle\Psi_1^\dagger \Psi_2$ in the scalar potential that leads to the $\Psi_1$ VEV as well as mixing among $\Psi_2$ and $\Psi_1$.

For a general $a \in \mathbb{R}$, the above particle content leads to a Lagrangian with conserved $U(1)_{L_\mu-L_\tau}\times U(1)_{B_1+B_2-2 B_3}$ symmetry. Both the $L_\mu-L_\tau$ symmetries and the $B_1+B_2-2 B_3$ are anomaly free and could be gauged, giving rise to two additional neutral gauge bosons $Z'$ and $Z''$ which can mix with the SM $Z$ boson~\cite{Heeck:2011md}. However, we only want to promote the linear combination of Eq.~\eqref{eq:generator} to a gauge symmetry in order to end up with a single $Z'$ that couples to quarks and leptons simultaneously. To remove the orthogonal global $U(1)''$ symmetry, generated by $Q'' = a (L_\mu-L_\tau)+ (B_1+B_2-2 B_3)$, we have to introduce couplings that connect e.g.~$\Phi_1$ and $\Phi_2$ non-trivially. This is only possible for $a\in \mathbb{Q}$ and in general requires the introduction of additional mediator fields. Let us sketch the simplest examples for $a$, using the fact that the phenomenologically interesting region will be $0<a < 1$.
\begin{itemize}
	\item For $a = 1/2$, the potential already allows for a term $\Phi_2^2 \Phi_1$ that breaks the orthogonal global $U(1)''$, so the particle content from above is sufficient to avoid Goldstone modes; a VEV for $\Phi_2$ will induce a VEV for $\Phi_1$.
\item For $a = 1/3$, it is the coupling $\Phi^3 \Phi_1$ that breaks the accidental global $U(1)''$ symmetry explicitly.
\item For $a = 1/4$, no dimension-4 operators can be written down with the given scalars that would break the global $U(1)''$. Therefore, one has to introduce a third singlet $\Phi_3$ with $Q'(\Phi_3) = 1/2$, which couples via $\Phi_3^2 \bar\Phi_1$ and $\Phi_2^2 \Phi_3$. A $\Phi_2$ VEV will induce a VEV for $\Phi_3$, which will induce a VEV for $\Phi_1$.
\item For $a=1/6$, we need $\Phi_3$ with $Q'(\Phi_3)= 1/3$ to couple $\Phi_3^3 \bar\Phi_1$ and $\Phi_2^2 \Phi_3$ similar to $a=1/4$.
\end{itemize}
The above considerations show that one can easily construct models for various values of $a \in \mathbb{Q}$. However, the details of this procedure will hardly make a difference in our discussion of the phenomenological effects of the doublet scalars and the $Z'$ in the following. Hence, we treat $a\in \mathbb{Q}$ as a free parameter but use $a=1/2$ and $a=1/3$ as benchmark values.

Similar to Refs.~\cite{Heeck:2014qea,Crivellin:2015mga} we consider the limit of small mixing between the heavy singlet scalars $\Phi_j$ and the two lighter doublets
\begin{align}
\Psi_j &\equiv \matrixx{\psi_j^+\\ (v_j + \psi_j^{0,{\rm R}} - i \psi_j^{0,{\rm I}})/\sqrt{2}} ,\quad j = 1,2\,.
\end{align}
Here, $\psi_{1,2}^{0,{\rm R}}$ and $\psi_{1,2}^{0,{\rm I}}$ correspond to the CP-even and the CP-odd components, respectively. 
For heavy unmixed singlet scalars, we end up with a restricted two-Higgs-doublet model (2HDM) potential of the form
\begin{align}
\begin{split}
V &\simeq  m_1^2 |\Psi_1|^2 + \tfrac{\lambda_1}{2} |\Psi_1|^4+ m_2^2 |\Psi_2|^2 + \tfrac{\lambda_2}{2} |\Psi_2|^4 \\
&\quad 	+ \lambda_{12} |\Psi_1|^2 |\Psi_2|^2 + \tilde\lambda_{12} |\Psi_1^\dagger \Psi_2|^2\\
&\quad + \left( m_{12}^2 \Psi_2^\dagger \Psi_1  +\hc \right),
\end{split}
\end{align}
where $m_{12}^2 \propto \mu \langle \Phi_2 \rangle$ is generated by the coupling $\mu \Phi_2 \Psi_1^\dagger \Psi_2$, which induces a small vacuum expectation value for $\Psi_1$~\cite{Heeck:2014qea}. We define $v \equiv \sqrt{v_1^2+v_2^2}\simeq 246\GeV$ and $\tan\beta =v_2/v_1$, which is medium to large in the region of interest. The neutral CP-even components $\psi_1^{0,{\rm R}}$ and $\psi_2^{0,{\rm R}}$ mix with an angle $\alpha$ in the usual 2HDM notation to give the mass eigenstates
\begin{align}
h &= \cos\alpha\, \psi_2^{{\rm R},0} - \sin\alpha\, \psi_1^{{\rm R},0}\,,\\
H &= \sin\alpha\, \psi_2^{{\rm R},0} + \cos\alpha\, \psi_1^{{\rm R},0}\,,
\end{align}
while the CP-odd components as well as the charged ones mix with $\beta$,
\begin{align}
A &= \cos\beta \psi^{0,{\rm I}}_2- \sin\beta \psi^{0,{\rm I}}_1\,,\\
H^- &= \cos\beta \psi^-_2- \sin\beta \psi^-_1\,.
\end{align}
Note that in the general 2HDM $-\pi/2<\alpha<\pi/2$ \cite{Branco:2011iw}. We will always assume that $h$ corresponds to the $125\GeV$ scalar discovered at the LHC~\cite{Aad:2012tfa,Chatrchyan:2012ufa}. Gauge bosons and leptons have standard type-I 2HDM couplings to the scalars (see for example Ref.~\cite{Branco:2011iw}).

\subsection{Quark masses and couplings}

The $U(1)'$-neutral scalar doublet $\Psi_2$ gives flavour diagonal mass terms for quarks and leptons, while $\Psi_1$ couples only off diagonally to quarks:
\begin{align}
{\cal L}_{{Y_q}}^{}\; &=  - {\bar Q_f}\left( {\xi _{fi}^u{{\tilde \Psi }_1} + Y_{fi}^u{{\tilde \Psi }_2}} \right){u_i} \\
&\quad - {\bar Q_f}\left( {\xi _{fi}^d{\Psi _1} + Y_{fi}^d{\Psi _2}} \right){d_i}\; + \hc
\label{eq:yukawas}
\end{align}
Here, $Q$ is the left-handed quark doublet, $u$ is the right-handed up quark and $d$ the right-handed down quark, while $i$ and $f$ label the three generations. We also define the doublets $\tilde \Psi_j \equiv i\sigma_2 \Psi_j^*$. The Yukawa couplings $Y^{u}$ and $Y^{d}$ of $\Psi_2$ are forced by our charge assignment to have the form
\begin{equation}
Y^q = \matrixx{
   {Y_{11}^q} & {Y_{12}^q} & 0  \\
   {Y_{21}^q} & {Y_{22}^q} & 0  \\
   0 & 0 & {Y_{33}^q}  } 
\end{equation}
(with $q=u,d$) and hence allow for the generation of the Cabibbo mixing connecting the first two generations, while the third generation is decoupled. The couplings $\xi^{q}$ are given by
\begin{equation}
\xi^u = \left( {\begin{array}{*{20}{c}}
   0 & 0 & 0  \\
   0 & 0 & 0  \\
   {{\xi _{tu}}} & {{\xi _{tc}}} & 0  \\
\end{array}} \right) ,\quad
\xi^d = \left( {\begin{array}{*{20}{c}}
   0 & 0 & {{\xi _{db}}}  \\
   0 & 0 & {{\xi _{sb}}}  \\
   0 & 0 & 0  \\
\end{array}} \right) 
\end{equation}
and lead to the small mixing of the first two generations with the top and bottom quarks after electroweak symmetry breaking.
The quark-mass matrices in the interaction eigenbasis are then given by
\begin{align}
m_u^{\rm EW}  &= \frac{v}{\sqrt{2}} \left(\sin\beta Y^u +\cos\beta \xi^u\right) \equiv U_L m_u^D U_R^\dagger\,,\\
m_d^{\rm EW}  &= \frac{v}{\sqrt{2}} \left(\sin\beta Y^d +\cos\beta \xi^d\right) \equiv D_L m_d^D D_R^\dagger\,,
\end{align}
related to the diagonal mass matrices in the physical basis
\begin{align}
m_u^D &= \diag (m_u, m_c, m_t)\,, \\
m_d^D &= \diag (m_d, m_s, m_b)
\end{align}
by the unitary matrices $U_{L,R}$ and $D_{L,R}$. The CKM matrix is then given by the misalignment of the left-handed up and down quark rotations as
\begin{equation}
	V\equiv U_L^\dagger D_L\,.
\end{equation}

The Lagrangian describing the couplings of quarks to the physical scalar fields is given by
\begin{widetext}
\begin{align}
\begin{split}
{\cal L} \supset \,&- \bar u \left(\frac{\cos\alpha}{v \sin\beta} m_u^D - \frac{\cos (\alpha-\beta)}{\sqrt{2}\sin\beta} \tilde\xi^u \right)P_R u \, h
- \bar d \left(\frac{\cos\alpha}{v \sin\beta} m_d^D-\frac{\cos(\alpha-\beta)}{\sqrt{2}\sin\beta} \tilde\xi^d \right)P_R d \, h \\
&- \bar u \left(\frac{\sin\alpha}{v \sin\beta} m_u^D - \frac{\sin (\alpha-\beta)}{\sqrt{2}\sin\beta} \tilde \xi^u \right)P_R u \, H
- \bar d \left(\frac{\sin\alpha}{v \sin\beta} m_d^D-\frac{\sin(\alpha-\beta)}{\sqrt{2}\sin\beta} \tilde \xi^d \right)P_R d \, H \\
&-i \bar u \left( \frac{m_u^D }{v \tan\beta}  - \frac{1}{\sqrt{2}\sin\beta} \tilde\xi^u \right) P_R u \, A +i \bar d\left( \frac{m_d^D }{v \tan\beta}  - \frac{1}{\sqrt{2}\sin\beta} \tilde \xi^d \right) P_R d \, A \\
& -\bar u \left[\left( \frac{\sqrt{2}}{v \tan\beta} m_u^D V  - \frac{1}{\sin\beta}  (\tilde\xi^u)^\dagger  V\right) P_L + \left( \frac{\sqrt{2} }{v \tan\beta} V m_d^D  - \frac{1}{\sin\beta} V  \tilde\xi^d \right) P_R \right] d\,H^+ \,,
\end{split}
\end{align}
\end{widetext}
where we omitted the hermitian conjugated terms and flavour indices. In addition we defined the non-diagonal coupling matrices
\begin{equation}
\tilde \xi^u= U_L^\dagger \xi^u U_R\,,\qquad
\tilde \xi^d=D_L^\dagger \xi^d D_R\,.
\end{equation}
The terms proportional to $m_q^D$ correspond to the usual type-I 2HDM-like couplings, while the terms involving $\tilde \xi^q$ induce flavour violation and will be regarded as small perturbations with respect to the type-I structure.

Since $\xi^u_{31}$ and $\xi^u_{32}$ correspond to right-handed rotations, the CKM matrix is (to a good approximation) given by $V \simeq D_L$, while the mixing angles in $D_R$ generated by $\xi^d_{13}$ and $\xi^d_{23}$ are suppressed to those in $D_L$ by $m_{d,s}/m_b$, and so $D_R \simeq \mathbb{I}$. A perturbative diagonalization of the quark mass matrices gives
\begin{equation}
{\xi _{db}} \simeq \frac{\sqrt{2}}{\cos\beta}\frac{{{m_b}}}{{{v}}}{V_{ub}}\,,\;\;{\xi _{sb}} \simeq  \frac{\sqrt{2}}{\cos\beta}\frac{{{m_b}}}{{{v}}} {V_{cb}}\,,
\end{equation}
in particular $\xi_{db}/\xi_{sb}\ll 1$. For medium to large values of $\tan\beta$, our model explains why $V_{ub}$ and $V_{cb}$ are much smaller than the Cabibbo angle as the contributions of $\xi _{sb}$ and $\xi_{db}$ to the mass matrix are suppressed by $\tan\beta$.\footnote{Demanding perturbatively small values for $\xi_{db,sb}$ yields a mild upper bound $\tan\beta < \mathcal{O}(10^3)$ due to the smallness of $V_{ub,cb}$.}
The non-diagonal matrix relevant for all the scalar couplings is then given by
\begin{align}
\tilde \xi^d &\simeq V^\dagger \xi^d \simeq 
\frac{\sqrt{2}}{\cos\beta} \frac{m_b}{v}
\matrixx{
0 & 0 & -V^*_{td} V_{tb}\\ 
0 & 0 & -V^*_{ts} V_{tb}\\
0 & 0 & 1-|V_{tb}|^2} \,.
\label{eq:approxcouplings}
\end{align}
The dominant flavour off-diagonal coupling is that to $sb$, while that to $db$ is suppressed by $V_{td}/V_{ts}\simeq 0.2$. Note that the flavour-violating couplings are entirely given in terms of the CKM matrix elements. Therefore, flavour violation in the down-quark sector induced by the scalars only involves the free parameters $\tan\beta$, $\alpha$ and the scalar masses, making our model highly predictive.

While the couplings $\xi^d_{23}$ and $\xi^d_{13}$ are fixed by the requirement that the measured CKM matrix is generated, the couplings $\xi^u_{31}$ and $\xi^u_{32}$ are free parameters generating top-quark flavor-changing effects not under investigation in this analysis. 
We therefore neglect their effect in the following and set $\xi^u = 0$. One should keep in mind that these couplings can in principle induce  decays such as $t\to h c$ and $t\to h u$. However, as we will see, the scalar-mixing angle $\alpha$ must be very small in order not to violate bounds from $B_s$--$\overline {B}_s$ mixing, rendering the $h$--$q$--$q$ couplings nearly SM-like. Therefore, also the effect in $t\to h c$ and $t\to h u$ is suppressed. 

The $Z'$ couplings to up quarks are vector-like in the limit $\xi^u = 0$ we consider in this paper
\begin{equation}
g'\, {{{\bar u}}}{\gamma ^\mu }\diag(-a/3, -a/3, 2a/3) {{u}}\, {Z'_\mu }\,.
\label{eq:up_current}
\end{equation}
Due to the rotation $D_L \simeq V$, the $Z'$ couplings to down quarks are chiral and given by
\begin{equation}
g'\left({{{\bar d}}_i}{\gamma ^\mu }{P_L}{{d}_j}{Z'_\mu }\Gamma_{ij}^{{d}L} + {{{\bar d}}}_i{\gamma ^\mu }{P_R}{{d}_j}{Z'_\mu }\Gamma_{ij}^{{d}R}\right) ,
\label{eq:down_current}
\end{equation}
with coupling matrices
\begin{align}
\Gamma^{dL} &\simeq a{\left( {\begin{array}{*{20}{c}}
   {\left| {{V_{td}}} \right|}^2 -\frac13 & {{V_{ts}}V_{td}^*} & {{V_{tb}}V_{td}^*}  \\
   {V_{td}^{}V_{ts}^*} & {{\left| {{V_{ts}}} \right|}^2-\frac13} & {{V_{tb}}V_{ts}^*}  \\
   {V_{td}^{}V_{tb}^*} & {V_{ts}^{}V_{tb}^*} & {{{\left| {{V_{tb}}} \right|}^2-\frac13}}  \\
\end{array}} \right)} \,,\\
\Gamma^{dR} &\simeq a{\left( {\begin{array}{*{20}{c}} -1/3 & 0 & 0  \\
   0 & -1/3 & 0  \\
   0 & 0 & 2/3  \\
\end{array}} \right)} \,.
\end{align}
Note that this reduces to $\diag(-a/3, -a/3, 2a/3)$ also for the left-handed couplings in the limit in which the CKM matrix is the unit matrix. As the right-handed couplings are to a good approximation flavour diagonal, this will ultimately lead to the desired hierarchy $|C_9^\prime| \ll |C_9|$ hinted at by global fits~\cite{Descotes-Genon:2013wba,Descotes-Genon:2013zva,Altmannshofer:2014rta}.
Note that an opposite $U(1)'$ charge for $\Psi_1$ would require the CKM matrix to be generated in the up sector, while right-handed down-quark $Z'$ couplings would be free parameters, rendering the left-handed down-quark $Z'$ couplings flavour conserving. As we will see in the next section, one cannot explain $R(K)$ and $B\to K^*\mu\mu$ in such a setup. Also note that the coupling matrix $\Gamma^{d L}$ is hermitian with complex off-diagonal entries. This will in principle lead to quark-dipole moments~\cite{Chiang:2006we}, which are, however, proportional to $\Gamma^L_{ij} \Gamma^{R\,*}_{ij}$ (with $i\neq j$) and hence vanishingly small in our scenario with diagonal $\Gamma^R$.

It is important to reiterate that our model has, by construction, suppressed flavour violation between the first two generations, which drastically softens constraints. Furthermore, the ``detached'' third quark generation is motivated by the observed smaller mixing angles.

\subsection{Lepton masses and couplings}
\label{sec:neutrinos}

Since we only consider $|a|<1$ in order to avoid stringent constraints from $\Delta F=2$ processes, it is not possible to couple the scalar doublet $\Psi_1$ with $U(1)'$ charge $a$ to leptons. The Dirac mass matrices for the leptons are hence generated by $\Psi_2$ only and are diagonal in flavour space due to the charge assignment of our $U(1)'$ symmetry. There is hence no charged-lepton flavor violation in this simple 2HDM. However, extending the model by a third doublet with $|Q'| = 2$ (Sec.~\ref{sec:3HDM}) can again lead to LFV and in particular give rise to the decay $h\to\mu\tau$, as pointed out in Refs.~\cite{Heeck:2014qea,Crivellin:2015mga}. For now, we consider the 2HDM with charged-lepton flavour conservation.

\begin{figure*}
\includegraphics[width=0.42\textwidth]{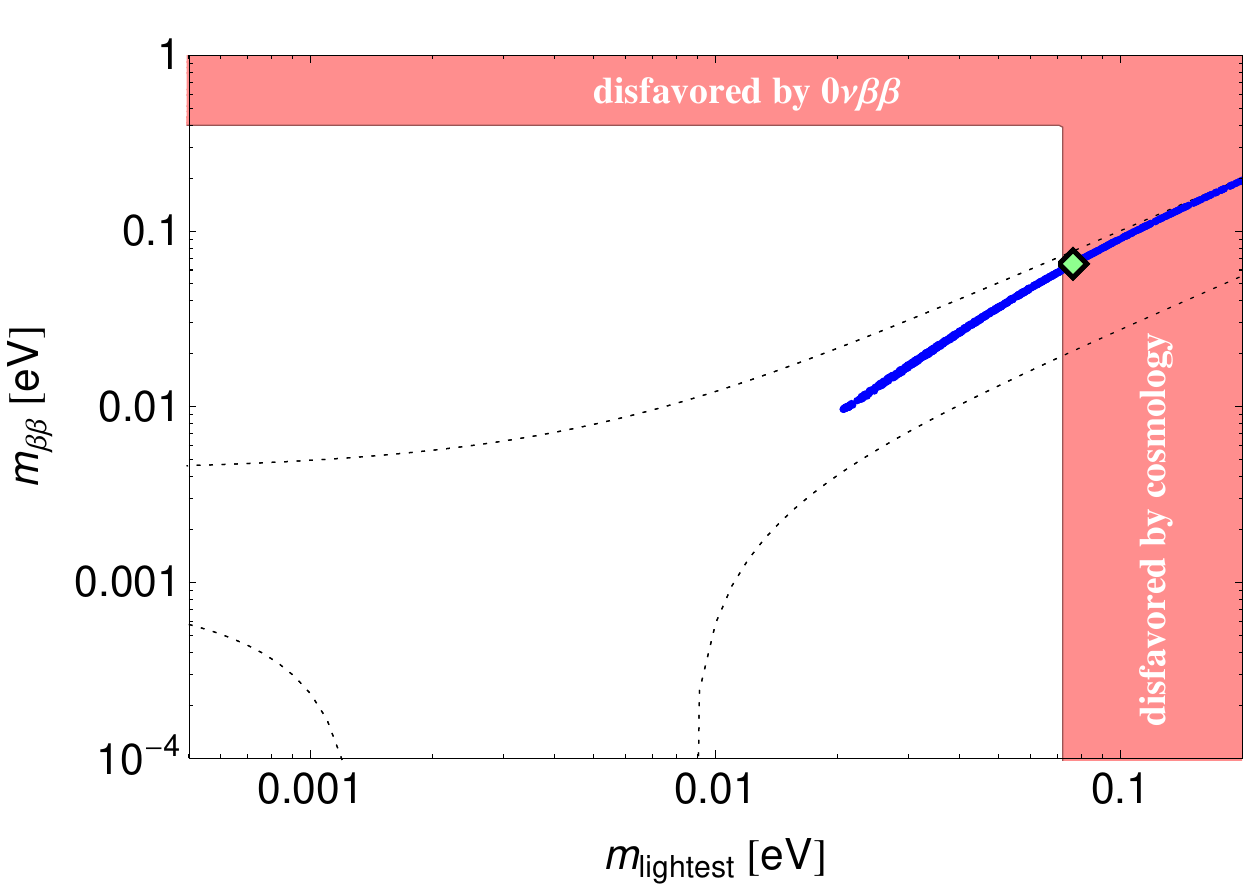} \hspace{3ex}
\includegraphics[width=0.42\textwidth]{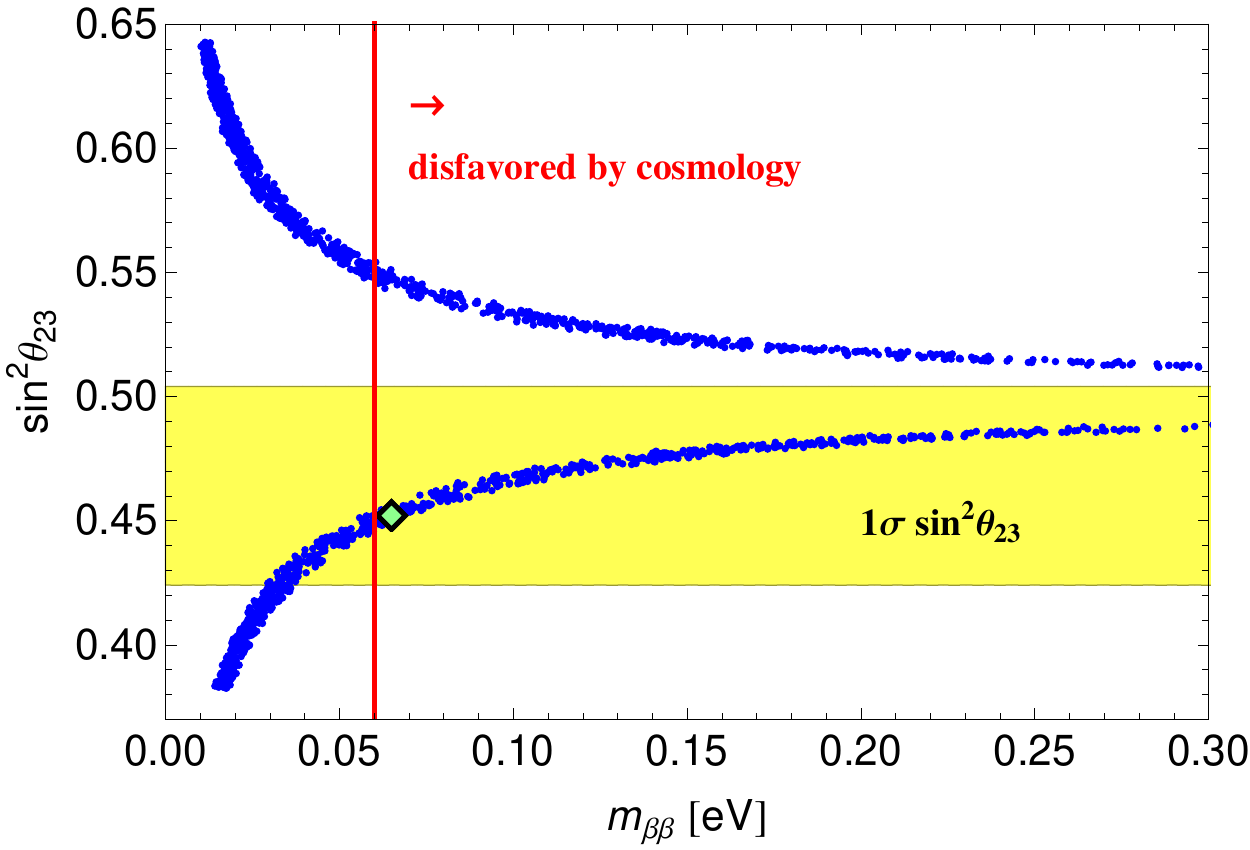}
\caption{Left: $m_{\beta\beta}$ vs.~lightest neutrino mass (in blue) as predicted by the two vanishing minors $(m_\nu^{-1})_{22} = 0 = (m_\nu^{-1})_{33}$ in combination with the $3\sigma$ values for $\theta_{ij}$ and $\Delta m_{ij}^2$ from Ref.~\cite{Gonzalez-Garcia:2014bfa}. The diamond marks the best-fit point. The dotted black lines show the standard allowed $3\sigma$ range for normal ordering. The red regions are disfavored by GERDA~\cite{Agostini:2013mzu} (conservative limit) and Planck~\cite{Ade:2013zuv,Planck:2015xua}.
Right: Correlation of $\sin^2 \theta_{23}$ with $m_{\beta\beta}$ due to the two vanishing minors. The yellow region denotes the $1\sigma$ range for $\theta_{23}$, and the diamond marks the best-fit point.}
\label{fig:neutrinos}
\end{figure*}

Neutrino masses arise via the seesaw mechanism \cite{Minkowski:1977sc} as $m_\nu \simeq - m_D m_R^{-1} m_D^T$ with diagonal Dirac mass matrix $m_D\propto \langle \Psi_2\rangle$ and the right-handed Majorana neutrino mass matrix given by~\cite{Araki:2012ip}
\begin{align}
m_R = \matrixx{M_1 & a_{12} \langle \Phi_1\rangle& a_{13} \langle \Phi_1\rangle \\  a_{12} \langle \Phi_1\rangle & 0 & M_2\\ a_{13} \langle \Phi_1\rangle & M_2 & 0} .
\label{eq:right-handed_mass}
\end{align}
As a consequence of our minimal singlet sector, two texture zeros in $m_R$ remain and propagate to $m_\nu$ as two vanishing minors~\cite{Araki:2012ip}: $(m_\nu^{-1})_{22} = 0 = (m_\nu^{-1})_{33}$. These zeros lead to relations among the neutrino mixing parameters that allow us to predict the unknown phases and masses given the measured neutrino mixing angles and the mass-squared differences~\cite{Lavoura:2004tu,Lashin:2007dm}. In our case we predict normal ordering, typically with quasi-degenerate masses. Delicate cancellations can, however, occur which lead to a normal hierarchy (see Fig.~\ref{fig:neutrinos}). With the best-fit values for $\Delta m_{ij}^2$ and $\theta_{ij}$ from the global fit of Ref.~\cite{Gonzalez-Garcia:2014bfa}, we predict the lightest neutrino mass to be $m_1 = 0.076\eV$, and hence $\sum_j m_j = 0.24\eV$, as well as $|\sin\delta_\mathrm{CP}| = 0.96$, and $m_{\beta\beta} \equiv |(m_\nu)_{ee}| = 0.065\eV$. Note that this prediction for $\delta_\mathrm{CP}$ agrees well with the preferred region from the global fit, $\delta_\mathrm{CP}/{}^\circ= 306^{+39}_{-70}$~\cite{Gonzalez-Garcia:2014bfa}. Using the $3\sigma$ ranges for mixing angles and $\Delta m_{ij}^2$, we obtain lower bounds on the neutrino mass observables
\begin{align}
\sum_j m_j \gtrsim 0.095 \eV \,, && 
m_{\beta\beta} \gtrsim 9.5 \meV\,.
\label{eq:neutrino_predictions}
\end{align}
The sum of neutrino masses $\sum_j m_j$ is large enough to have a cosmological impact~\cite{Abazajian:2013oma}; in fact, Planck already constrains $\sum m < 0.23 \eV$ at $95\%$~C.L.~\cite{Ade:2013zuv,Planck:2015xua}, which, however, depends strongly on the combination of data sets. The effective neutrino mass $m_{\beta\beta}$ relevant for neutrinoless double-beta decay is also in the reach of future experiments (see Ref.~\cite{Bilenky:2014uka} for a recent review).
There is a strong correlation between the CP phase and the atmospheric mixing angle, approximately given by
\begin{align}
\delta_\mathrm{CP} \simeq 
\begin{cases} 90^\circ + 360^\circ (\sin^2\theta_{23} - 1/2) \,, \mathrm{or}	\\
 270^\circ - 360^\circ (\sin^2\theta_{23} - 1/2)\,.
\end{cases}
\end{align}
For a more extensive discussion of this neutrino structure, we refer to Refs.~\cite{Lavoura:2004tu,Lashin:2007dm}.
Despite the lack of charged lepton flavour violation, our simple 2HDM is hence testable in the lepton sector as well.

Let us come to the scalar--lepton interactions. Since $\Psi_1$ has no couplings to leptons, the couplings of $h$, $H$, $A$, and $H^+$ are given simply by those of a type-I 2HDM, i.e.
\begin{align}
\begin{split}
{\cal L} \supset \,&- \bar e \left(\frac{\cos\alpha}{v \sin\beta} m_e^D\right) e \, h - \bar e \left(\frac{\sin\alpha}{v \sin\beta} m_e^D\right) e \, H \\
& -i \bar e\left( \frac{m_e^D }{v \tan\beta}  \right) \gamma_5 e \, A \\
& -\left[\bar \nu \left( \frac{\sqrt{2} }{v \tan\beta} U^\dagger m_e^D  \right) P_R  e\,H^+ +\hc\right] ,
\end{split}
\end{align}
with the Pontecorvo--Maki--Nakagawa--Sakata mixing matrix $U$ and $m_e^D \equiv \diag (m_e, m_\mu, m_\tau)$. 
Here, we ignored the couplings to the right-handed neutrinos, assuming them to be very heavy.
For the charged-lepton couplings to $Z'$, we have the usual vectorial $L_\mu-L_\tau$ couplings
\begin{equation}
g'\left({{{\bar e}}_i}{\gamma ^\mu }{P_L}{{e}_j}{Z'_\mu }\Gamma_{ij}^{{e}L} + {{{\bar e}}}_i{\gamma ^\mu }{P_R}{{e}_j}{Z'_\mu }\Gamma_{ij}^{{e}R}\right) ,
\label{eq:lepton_current}
\end{equation}
with
\begin{equation}
\Gamma _{fi}^{e L} = \Gamma _{fi}^{e R} = \left( {\begin{array}{*{20}{c}}
   0 & 0 & 0  \\
   0 & 1 & 0  \\
   0 & 0 & { - 1}  \\
\end{array}} \right) ,
\end{equation}
which are flavour-conserving in this minimal 2HDM (lepton-flavour violation will arise in the 3HDM of Sec.~\ref{sec:3HDM}). 

\subsection{Gauge boson sector}
\label{sec:gauge}

The $Z'$ mass in case of just two singlet scalars (see Eq.~\eqref{eq:singlets}) is given in the limit of interest $\langle\Phi_j\rangle \gg \langle\Psi_i\rangle $ by
\begin{align}
m_{Z'}^2 \simeq 2 {g'}^2\left( \langle\Phi_1\rangle^2 + a^2 \langle\Phi_2\rangle^2 \right) .
\end{align}
Ignoring kinetic mixing between $U(1)'$ and hypercharge $U(1)_Y$, a $Z$--$Z'$ mixing angle $\theta_{Z Z'}$~\cite{Langacker:2008yv} is nevertheless induced by the VEV of $\Psi_1$~\cite{Heeck:2014qea},
\begin{align}
\begin{split}
 g' \theta_{Z Z'} &\simeq -\frac{a g_1  v^2 \cos^2\beta}{2 m_{Z'}^2/g'^2}\\
 &\simeq  -2\times 10^{-4}\,a\,\left(\frac{10}{\tan\beta}\right)^2 \left(\frac{\TeV}{m_{Z'}/g'}\right)^2 .
\end{split}
\label{eq:ZZpmixing}
\end{align}
The gauge eigenstates $Z$ and $Z'$ can then be expressed in terms of the mass eigenstates $Z_1$ and $Z_2$ as 
\begin{align}
Z &\simeq Z_1 -\theta_{Z Z'} Z_2\,, \\
Z' &\simeq Z_2 +\theta_{Z Z'} Z_1\,,
\end{align}
which couple to
\begin{align}
-{\cal L}\supset  \left( g_1 j_{1,\mu} + g' \theta_{Z Z'} j'_\mu\right) Z_1^\mu+ \left( g' j'_\mu -  g_1  \theta_{Z Z'}j_{1,\mu} \right) Z_2^\mu \,,
\end{align}
where $g_1 \equiv g/c_W = e/s_W c_W$ (with $s_W = \sin\theta_W$ etc.~for the weak mixing angle $\theta_W$) and the SM neutral current~\cite{Langacker:2008yv}
\begin{align}
j_1^\mu = \sum_f \bar f  \gamma^\mu \left[ (t_3(f)-s_W^2 Q(f)) P_L - s_W^2 Q(f) P_R\right]f \,.
\end{align}
The $U(1)'$ current $j'_\mu$ is given in Eqs.~\eqref{eq:up_current}, \eqref{eq:down_current}, and~\eqref{eq:lepton_current}.

In the lepton sector, this $Z$--$Z'$ mixing leads to small vector-coupling shifts of the light $Z_1$ (i.e. the SM $Z$ studied at LEP) to muons and taus,
\begin{align}
g_V^Z (\mu\mu, \tau\tau) \simeq -1/2 + 2 s_W^2 \pm g' \theta_{Z Z'}/g_1 \,,
\end{align}
and thus ultimately to lepton-flavour non-universality in the $Z$ couplings~\cite{Heeck:2011wj}. Simply demanding that the additional contribution does not exceed three times the $1\sigma$ error ($g_V^Z(\tau\tau) = -0.0366 \pm 0.0010$) gives
\begin{align}
|g' \theta_{Z Z'} |\lesssim g_1\, 0.003 \simeq 2\times 10^{-3} \,,
\end{align}
easily satisfied for the parameters under investigation (see Eq.~\eqref{eq:ZZpmixing}). 

The $Z$ also inherits the $Z'$ quark couplings and thus contributes to flavor violation in the quark sector. Its contribution will be enhanced by its smaller mass but suppressed by $\theta_{Z Z'}$. For TeV scale $Z'$, the mass-enhancement cannot overcome the $Z$--$Z'$ mixing suppression, so we ignore the flavour-changing $Z$ couplings in the following.

Since the $Z$--$Z'$ mixing angle is small, we will continue to denote the new gauge boson by $Z'$ in the following.
As we assume the three right-handed neutrinos to be heavier than the $Z'$, the invisible branching ratio is set by the active neutrinos,
\begin{align}
{\rm Br}[Z' \to {\rm inv}] &= \frac{\sum_\nu\Gamma [Z'\to \nu\nu]}{\sum_f\Gamma [Z'\to ff]}\simeq \frac{1}{3 + 4 a^2} \,,
\end{align}
assuming $m_t \ll m_{Z'} < 2 m_{\nu_R}$. In the same limit, we have
\begin{align}
{\rm Br}[Z' \to \mu\mu] \simeq {\rm Br}[Z' \to \tau\tau] \simeq \frac{1}{3 + 4 a^2} \,,
\end{align}
while the branching ratio into two up quarks (same for $d$, $s$, $c$) is $a^2/3$ times the above, that into top quarks (same for bottom) is $4 a^2/3$ times the above.

\section{Flavour observables and LHC constraints}
\label{flavour}

We will now investigate the relevant bounds on our model, i.e.~$B$-meson decays, $\Delta F=2$ processes, neutrino trident production, and direct LHC searches. We will not go into details about the standard phenomenology of 2HDMs of type I (see Refs.~\cite{Dumont:2014wha,Dumont:2014kna} for a recent evaluation) but rather discuss constraints arising from the deviations from the type-I structure due to the additional flavour violation.

\subsection{\texorpdfstring{\BKs and $B\to K\mu^+\mu^-/B\to K e^+ e^-$}{B-meson decays}}

Concerning leptonic $B$ decays, both \BKs and $R(K)$ are sensitive to the Wilson coefficients $C^{(\prime)\mu\mu}_9$ and $C^{(\prime)\mu\mu}_{10}$ incorporated in the effective Hamiltonian,
\begin{equation}
 {\cal H}_\mathrm{eff} =  - \frac{{4{G_F}}}{{\sqrt 2 }}{V_{tb}}V_{ts}^*\left( \sum_{j=9,10} C^{\ell\ell}_j O^{\ell\ell}_j + C^{\prime\ell\ell}_j O^{\prime\ell\ell}_j \right) + \hc\,,
\end{equation}
with
\begin{align}
\begin{split}
 {O_9^{\ell\ell}}       &= \frac{\alpha_\mathrm{EM}}{{4\pi}} \left[\bar s{\gamma^\mu }{P_L}b\right] \left[\bar \ell {\gamma_\mu }\ell\right]  , \\
 O_9^{\prime\ell\ell}    &= \frac{\alpha_\mathrm{EM}}{{4\pi}}  \left[\bar s{\gamma^\mu }{P_R}b\right] \left[\bar \ell {\gamma_\mu }\ell \right] ,\\
 {O_{10}^{\ell\ell}}    &= \frac{\alpha_\mathrm{EM}}{{4\pi}}  \left[\bar s{\gamma^\mu }{P_L}b\right] \left[\bar \ell {\gamma_\mu }{\gamma^5}\ell \right] , \\
 O_{10}^{\prime\ell\ell} &= \frac{\alpha_\mathrm{EM}}{{4\pi}}  \left[\bar s{\gamma^\mu }{P_R}b \right]\left[\bar \ell {\gamma_\mu }{\gamma^5}\ell \right] ,
\end{split}
\end{align}
with $\ell\in\{e,\mu,\tau\}$. In our model we generate the two coefficients $C_9^{\mu\mu} = - C_9^{\tau\tau}$ with
\begin{equation}
{C^{\mu\mu}_9} \simeq \frac{-{ {{g'}^2}}}{{\sqrt 2 m_{Z'}^2}}\frac{\pi }{\alpha_\mathrm{EM} }\frac{1}{G_F}a\simeq -\left(\frac{a}{1/3}\right) \left(\frac{3\TeV}{m_{Z'}/g'}\right)^2 ,
\label{eq:C9}
\end{equation}
while
\begin{align}
C_9^{ee}= C_9^{\prime \ell\ell} = C_{10}^{\ell\ell} = C_{10}^{\prime \ell\ell} = 0 \,.
\end{align}
Note that $C_9^{\mu\mu}$ is real in our model but could have either sign depending on $a$.
As already noted in Refs.~\cite{Descotes-Genon:2013wba,Descotes-Genon:2013zva}, $C^{\mu\mu}_9<0$ and $C^{\prime\mu\mu}_9=0$ gives a good fit to data. Using the global fit of Ref.~\cite{Altmannshofer:2014rta} we see that at ($1\,\sigma$) $2\,\sigma$ level 
\begin{equation}
 -0.60\, (-0.95) \geq  C_9^{\mu\mu}	\geq(-1.65)\, -2.00 \,,
\label{eq:C9fit}
\end{equation}
with a best-fit value $C_9^{\mu\mu} \simeq -1.3$.
Interestingly, the regions for $C_9^{\mu\mu}$ required by $R(K)$ and \BKs lie approximately in the same region. 
A $Z'$ explanation of $R(K)$ and \BKs has also been considered in Ref.~\cite{Altmannshofer:2014rta} as an effective model, with similar constraints as presented here.

Putting everything together, we find the $2\,\sigma$ preferred region for our model from $R(K)$ and \BKs
\begin{equation}
2.3 \dfrac{m_{Z'}^2}{(10\,{\rm TeV})^2}\leq a g'^2\leq 7.7	\dfrac{m_{Z'}^2}{(10\,{\rm TeV})^2}\,,
\label{eq:C9bound}
\end{equation}
with best-fit value $m_{Z'}/\sqrt{a}g' =  4.5\TeV$.
In the following we discuss additional constraints on $a$, $g'$ and $m_{Z'}$ that are of relevance to our resolution of the LHCb anomalies.

\subsection{\texorpdfstring{$\Delta F=2$ processes}{Delta F = 2 processes}}

The $Z'$ and scalars have flavour non-diagonal couplings, generating tree-level contributions to $\Delta F=2$ processes. Considering for definiteness $B_s$--$\bar B_s$ mixing ($B_d$--$\bar B_d$ and $K$--$\bar K$ follow by simple replacements of indices), the relevant effective Hamiltonian is conventionally written as (see for example Refs.~\cite{Ciuchini:1997bw,Crivellin:2013wna})
\begin{equation}
{\cal H}_\mathrm{eff}^{\Delta F = 2} = \sum\limits_{j = 1}^5 {{C_j}} {\mkern
  1mu} {O_j} + \sum\limits_{j = 1}^3 {{C_j'}} {\mkern 1mu} {O_j'} +\hc\,,
\end{equation}
with
\begin{align}
\begin{split}
 {O_1}  &= \left[ {{{\bar s}_\alpha }{\gamma ^\mu }{P_L}{b_\alpha }} \right]\left[ {{{\bar s}_\beta }{\gamma ^\mu }{P_L}{b_\beta }} \right],  \\ 
 {O_2} &= \left[ {{{\bar s}_\alpha }{P_L}{b_\alpha }} \right]{\mkern 1mu} \left[ {{{\bar s}_\beta }{P_L}{b_\beta }} \right] ,  \\
 {O_3}  &= \left[ {{{\bar s}_\alpha }{P_L}{b_\beta }} \right]\left[ {{{\bar s}_\beta }{P_L}{b_\alpha }} \right] ,\\ 
 {O_4}  &= \left[ {{{\bar s}_\alpha }{P_L}{b_\alpha }} \right]\left[ {{{\bar s}_\beta }{P_R}{b_\beta }} \right] ,  \\
 {O_5}  &= \left[ {{{\bar s}_\alpha }{P_L}{b_\beta }} \right]{\mkern 1mu} \left[ {{{\bar s}_\beta }{P_R}{b_\alpha }} \right] . 
\end{split}
\end{align}
Here, $\alpha$ and $\beta$ are color indices, and the primed operators are obtained by the exchange $L\leftrightarrow R$.

In our model we find the contributions (using the expression for $C^\prime_2$ given in Ref.~\cite{Crivellin:2013wna})
\begin{align}
{C_1} = \frac{g'^2}{{2m_{Z'}^2}}{{\left( {\Gamma _{sb}^{d L}} \right)}^2}\,, && C^\prime_2 = \sum\limits_{\eta = h,H,A} {\frac{{ - 1}}{{2m_\eta^2}}{\mkern 1mu} {{\left( {\Gamma _{sb}^\eta} \right)}^2}} ,
\end{align}
while all other Wilson coefficients are zero.
It is important to note that the $Z'$ contribution to the $\Delta F = 2$ processes scales like $a^2 g'^2/m_{Z'}^2$, while the contribution to leptonic decays, such as $B\to K \mu\mu$, are proportional to $a g'^2/m_{Z'}^2$. This makes it possible to evade $\Delta F = 2$ constraints by choosing $a \ll 1$, while still accommodating the LHCb anomalies. As we will see below, $a\lesssim 1$ is actually already sufficient for this purpose.

The Wilson coefficients $C_j$ enter physical observables, i.e.~mass differences and CP asymmetries, via the calculation of matrix elements involving decay constants and bag factors calculated with lattice QCD (see for example Ref.~\cite{Aoki:2013ldr} for a review of recent lattice values). In addition, the QCD renormalization group effects must be taken into account. For this we use the next-to-leading-order equations calculated in Refs.~\cite{Ciuchini:1997bw,Buras:2000if}.

\begin{figure}
\includegraphics[width=0.42\textwidth]{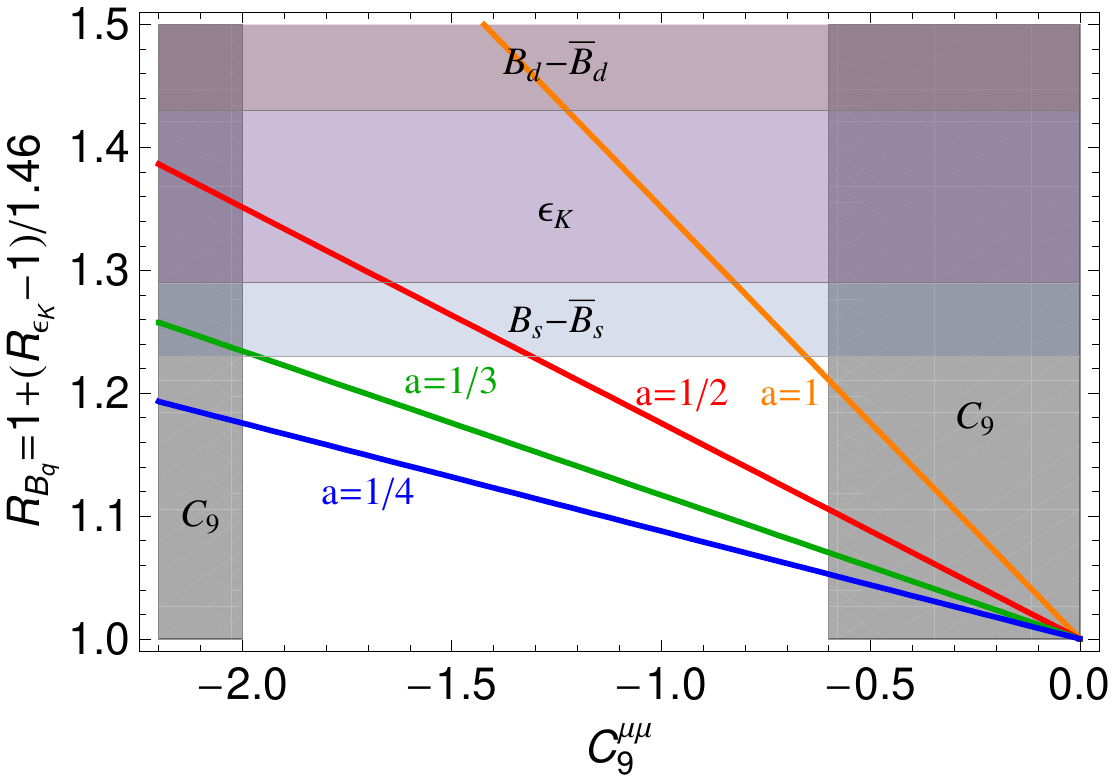}
\caption{$R_{B_q}$ and $R_{\epsilon_K}$ as a function of $C_9^{\mu\mu}$ for different values of $a$ taking into account the $Z'$ contribution only. The horizontal gray regions are excluded by $\Delta m_{B_q}$ and $\epsilon_K$, while the vertical ones are excluded by $B\to K^*\mu^+\mu^-$ and $R(K)$. We used $m_{Z'}=3\TeV$ for the renormalization group for the $\Delta F=2$ processes. Note that the dependence on $m_{Z'}$ is therefore only logarithmic.}
\label{fig:C9_RBq_RepsilonK}
\end{figure}

On the experimental side, the central values of $\Delta m_{B_s}$ and $\Delta m_{B_d}$ are slightly above the SM prediction, and the same is true for $\epsilon_K$ extracted from $K$--$\bar K$ mixing. This is interesting since our model predicts necessarily constructive interference of the $Z'$ contribution with the SM in all three observables. For our numerical values, we use the 95$\%$~C.L.~results of the UTfit collaboration \cite{UTfit,Bona:2006sa,Bona:2007vi}:
\begin{align}
0.76<R_{B_d}&=\Delta m_{B_d}/\Delta m_{B_d}^{\rm SM}<1.43\,,\\
0.90<R_{B_s}&=\Delta m_{B_s}/\Delta m_{B_s}^{\rm SM}<1.23\,,\\
0.77<R_{\epsilon_K}&=\epsilon_K/\epsilon_K^{\rm SM}<1.41\,.
\end{align}
Similar results are obtained by the CKMfitter collaboration~\cite{Charles:2004jd}.

The strongest constraints come from $B_s$ mixing, as can be seen in Fig.~\ref{fig:C9_RBq_RepsilonK}, but can easily be evaded for $a<1$ even for the large $|C_9^{\mu\mu}|\sim 1$ required to explain the $B$-meson anomalies. Similar but weaker bounds hold for the other $\Delta F=2$ processes.
From $B_s$--$\overline{B}_s$ mixing we get the approximate $95\%$~C.L.~bound
	\begin{equation}
	a^2g'^2\leq2.6\dfrac{m_{Z'}^2}{(10\,{\rm TeV})^2} \,.	
	\end{equation}
Note that this bound is obtained for $m_{Z'}=10\TeV$ and only scales approximately like $m_{Z'}^2$ due to the additional logarithmic dependence on the mass from the renormalization group.

\begin{figure}
\includegraphics[width=0.42\textwidth]{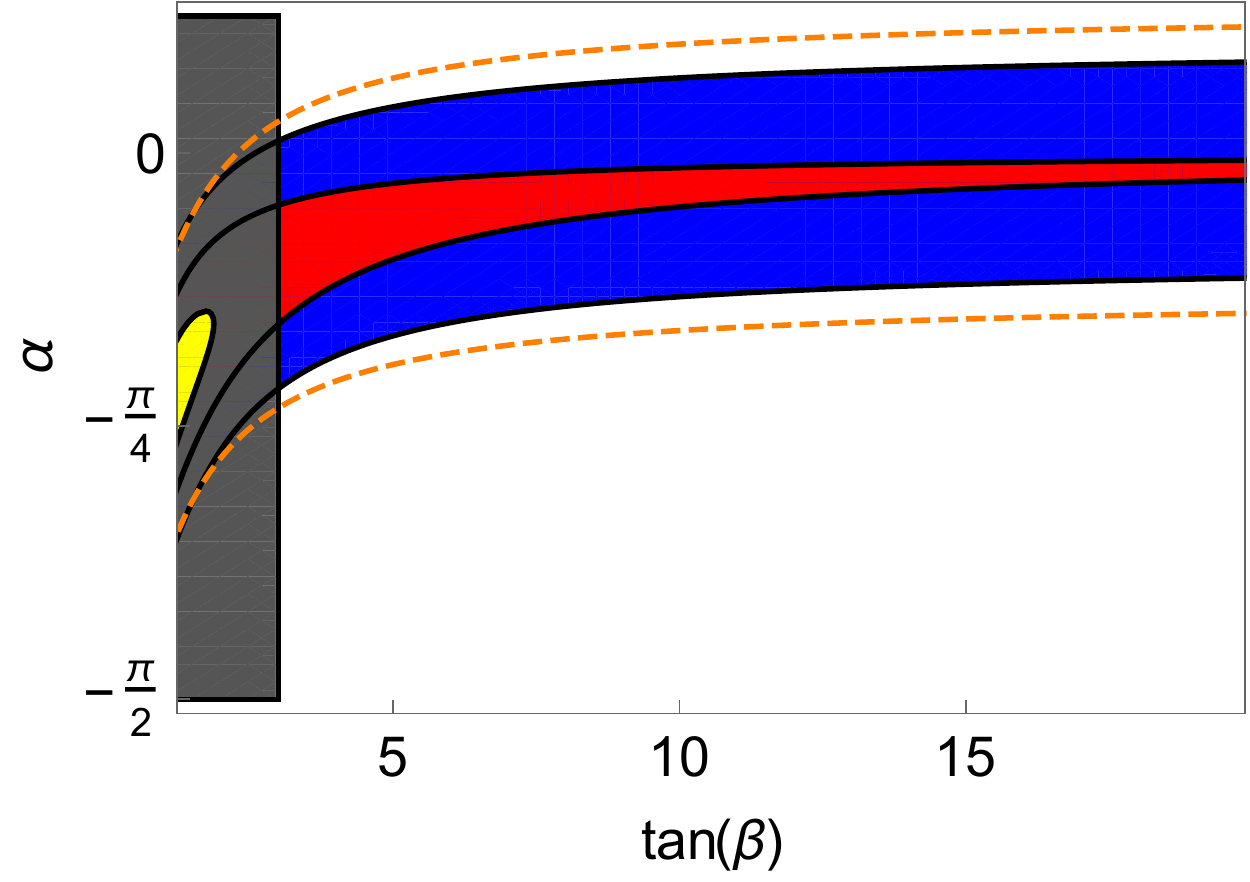}
\caption{Allowed regions in the $\tan\beta$--$\alpha$ plane assuming that $C_9^{\mu\mu}$ is reproduced by the $Z'$ contribution within $2\,\sigma$ for $a=1/3$ for $m_H=300\GeV$ and $m_A=350\GeV$ (yellow), $m_A=300\GeV$ (red) and $m_A=250\GeV$ (blue). The gray region is excluded by $b\to s\gamma$~\cite{Chen:2013kt}. Dashed lines indicate $\cos (\alpha-\beta) = \pm 0.4$ for reference.}
\label{fig:tanbeta-alpha}
\end{figure}

The flavour-changing neutral scalar couplings also affect $\Delta m_{B_s}$ (and also $\Delta m_{B_q}$ and $\Delta m_{K}$). Here, the $H,A$ contributions decouple as $1/m_{H,A}^2$, while the $h$ contributions vanishes for $\alpha-\beta=\pi/2$.
For large values of $\tan\beta$ and small $\alpha$ we have approximately
\begin{align}
\begin{split}
	\dfrac{\Delta m_{B_s}^{h,H,A}}{\Delta m_{B_s}^{\rm SM}}&\simeq 0.12 \cos^2(\alpha-\beta) \tan^2\beta \\
	&\quad+ 0.19  \tan^2\beta \left(\dfrac{(200\,{\rm GeV})^2}{m_{H}^2}-\dfrac{(200\,{\rm GeV})^2}{m_A^2}\right) .
	\end{split}
	\label{eq:DeltaMB_from_scalars}
\end{align}
The allowed regions in the $\tan\beta-\alpha$ plane are shown in the right plot of Fig.~\ref{fig:tanbeta-alpha} assuming that the $Z'$ generates the central value of $C_9^{\mu\mu}$ for $a=1/3$. One can see that the bounds are weaker if $m_A<m_H$ and stronger for $m_A>m_H$ due to the negative $A$ contribution in Eq.~\eqref{eq:DeltaMB_from_scalars}.

\subsection{Neutrino trident production}
\label{sec:NTP}

The most stringent bound on flavour-diagonal $Z'$ couplings to muons only (i.e.~no quark or electron couplings) arises from neutrino trident production (NTP) $\nu_\mu N \to \nu_\mu N \mu^+\mu^-$~\cite{Altmannshofer:2014cfa, Altmannshofer:2014pba}:
\begin{equation}
	\frac{\sigma_{\rm NTP}}{\sigma_{\rm NTP}^{\rm SM}}\simeq \frac{1+\left(1+4s_W^2+8\frac{g'^2}{m_{Z^\prime}^2}\frac{m_W^2}{g_2^2}\right)^2}{1+\left(1+4s_W^2\right)^2} \,.
\end{equation}
Taking only the CCFR data~\cite{Mishra:1991bv}, we find roughly $m_{Z'}/g' \gtrsim 550 \GeV$ at $95\%$~C.L.~for a heavy $Z'$. This cuts slightly into the parameter space allowed by $B_s$ mixing and $C_9$ but is only relevant for $a\ll 1$ (see Fig.~\ref{fig:VEVvsA}).

\subsection{Direct LHC searches}
\label{sec:LHC}

\begin{figure}[t]
\includegraphics[width=0.47\textwidth]{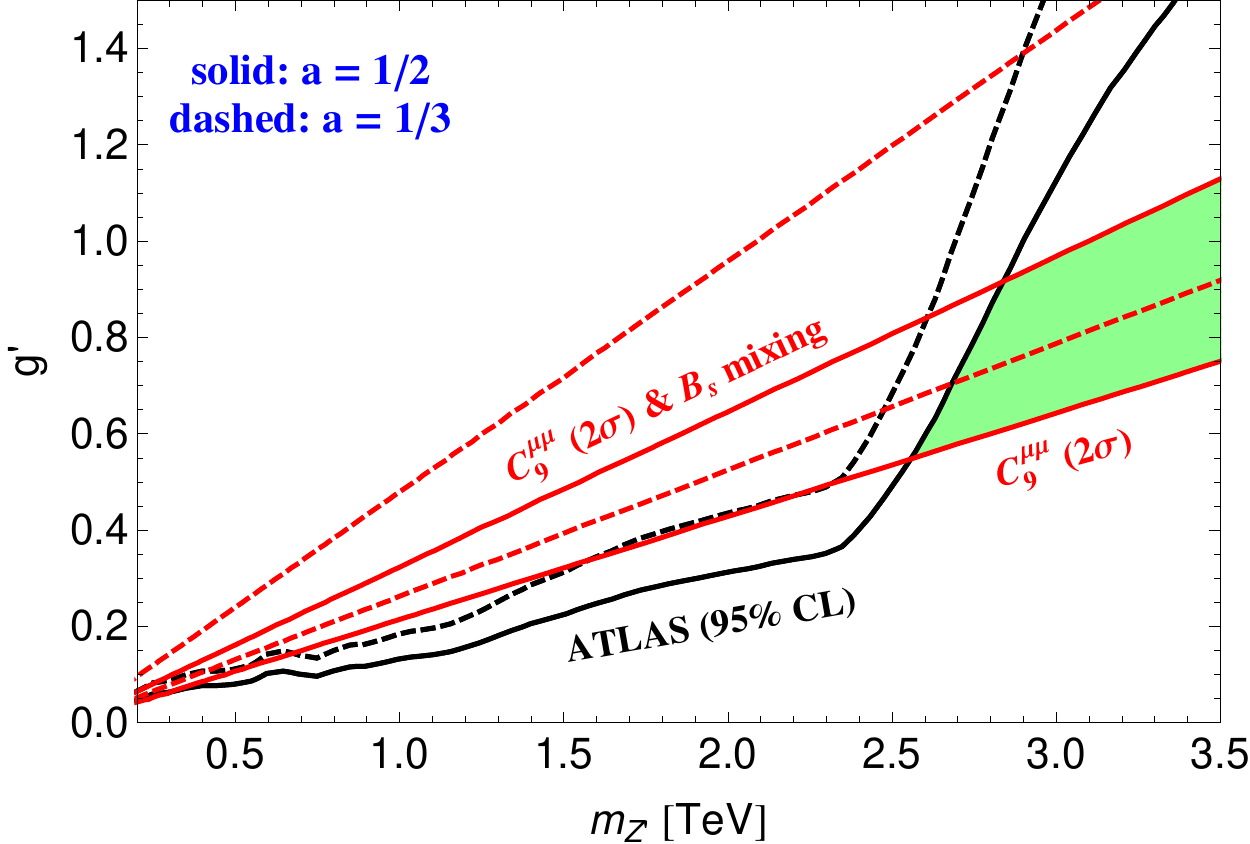}
\caption{Limits on $q\overline{q}\to Z' \to \mu\overline{\mu}$ from ATLAS~\cite{Aad:2014cka} (black, allowed region down right) and the $2\sigma$ limits on $C_9^{\mu\mu}$ to accommodate \BKs and $B\to K\mu^+\mu^-/B\to K e^+ e^-$ (red, allowed regions inside the cone). Solid (dashed) lines are for $a=1/2$ ($a=1/3$). For $a =1/2$, the green shaded region is allowed (similar for $a= 1/3$ using the dashed bounds).}
\label{ATLAS_limits}
\end{figure}

Since (unlike in Refs.~\cite{Altmannshofer:2014cfa,Crivellin:2015mga}) we have (potentially sizable) $Z'$ couplings to the first-generation quarks, our model is constrained by LHC searches for $pp \to Z' \to \mu\mu$. 
With vectorial $Z'$ couplings, universal in the first four quark generations, our model is closely related to $U(1)_{B-L}$ models~\cite{Heeck:2014zfa}, for which dedicated analyses exist. Working in the narrow-width approximation, the relevant quantity for collider searches is the Drell--Yan production cross section times branching ratio into muons,
\begin{align}
\frac{\sigma (pp\to Z') {\rm BR} [Z'\to\mu\mu]}{\sigma (pp\to Z_{B-L}') {\rm BR} [Z_{B-L}'\to\mu\mu]} \simeq \frac{a^2 g'^2}{g_{B-L}^2}\frac{1/(3+4 a^2)}{2/13} \,,
\end{align}
valid for $m_t \ll m_{Z'}< 2m_{\nu_R}$. One can therefore simply rescale the $B-L$ limits of ATLAS's $\sqrt{s}=8\TeV$ analysis of di-muon resonance searches (see auxiliary figures of Ref.~\cite{Aad:2014cka}), resulting in Fig.~\ref{ATLAS_limits}.
Contrary to low-energy observables -- which only depend on the ratio  $g'^2/m_{Z'}^2$ -- the LHC probes on-shell $Z'$. This leads to a complicated dependence on $m_{Z'}$, since the production cross section involves parton-density functions. As a result, the production cross section, and therefore the sensitivity, decreases strongly once $m_{Z'}$ approaches the maximal available energy.

For $a\ll 1$ (i.e.~the leptophilic case), the Drell--Yan production becomes negligible, and our model again resembles the standard $L_\mu-L_\tau$ models. In this case, the $Z'$ production at colliders goes through $pp \to \mu\mu Z' \to 4\mu$ (or $3\mu$ plus missing energy), where the $Z'$ is radiated off a final state lepton~\cite{Ma:2001md}. For $m_{Z'} > m_Z$, the LHC constraints are currently weaker than NTP~\cite{Altmannshofer:2014cfa} but will become competitive with higher luminosities~\cite{Harigaya:2013twa,delAguila:2014soa,Bell:2014tta}.

Finally, for $Z'$ masses above the on-shell threshold, one can obtain limits from searches for contact interactions, in our case
\begin{align}
{\cal L}\, \supset \,\frac{a}{3}\frac{g'^2}{m_{Z'}^2} \ \overline{q}\gamma^\alpha q\ \overline{\mu} \gamma_\alpha \mu \,, && \mathrm{ with } &&\, q\in \{u,d,s,c\}\,.
\end{align}
For positive $a$, the strongest limit from ATLAS is on the operator $\overline{q}\gamma^\alpha P_L q\ \overline{\mu} \gamma_\alpha P_R \mu$, providing a $95\%$~C.L.~limit of~\cite{Aad:2014wca}
\begin{align}
{m_{Z'}}/{g'} > 1.4\TeV \, \sqrt{{a}/{(1/3)}} \,,
\end{align}
which is weaker than the bounds from $C_9$ (Eq.~\eqref{eq:C9bound}).

\subsection{Discussion}
\label{pheno}

\begin{figure}
\includegraphics[width=0.45\textwidth]{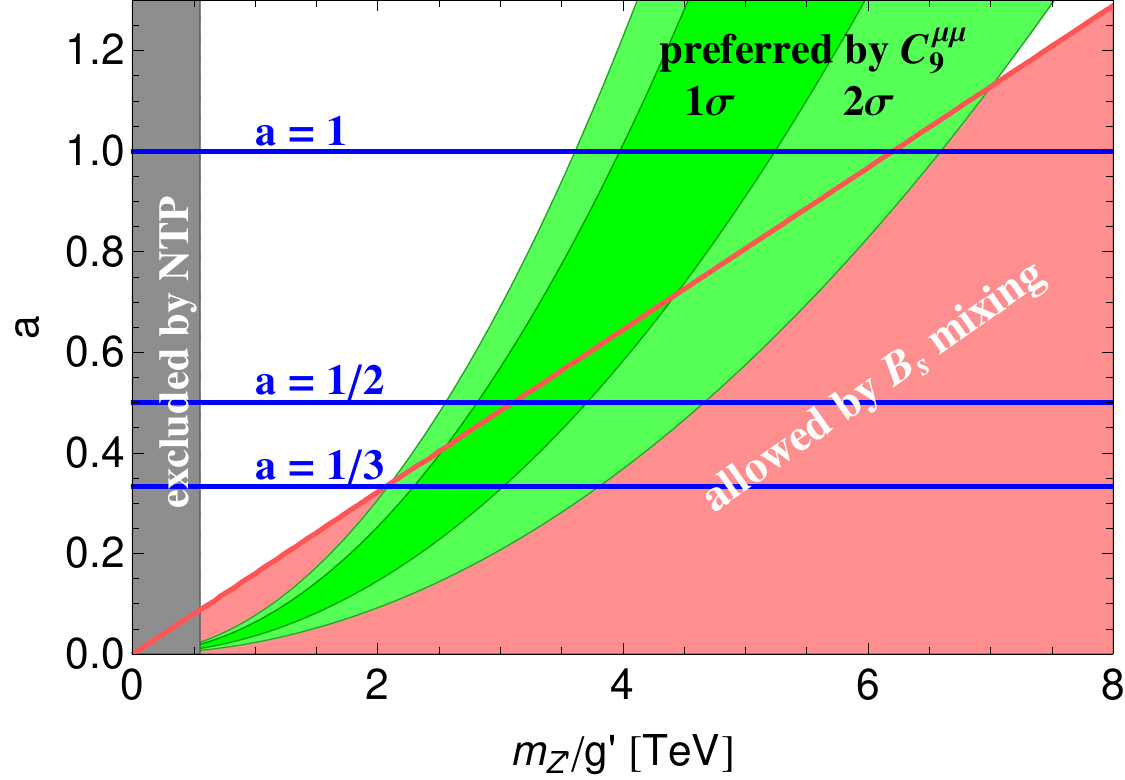}
\caption{Limits on $m_{Z'}/g'$ vs.~$a$ from NTP (gray), $B_s$--$\overline{B}_s$ mixing (red), and $C_9^{\mu\mu}$ (green). The horizontal lines indicate some values of interest: $a=1$, $1/2$, and $1/3$. Not shown are LHC limits (see Fig.~\ref{ATLAS_limits}).}
\label{fig:VEVvsA}
\end{figure}

The relevant low-energy constraints are collected in Fig.~\ref{fig:VEVvsA}.
If we want to explain \BKs and $B\to K\mu^+\mu^-/B\to K e^+ e^-$ within $2\sigma$ ($1\sigma$), we need $a< 1.13$ ($0.71$) to avoid stringent $B_s$--$\overline{B}_s$ mixing constraints (taking into account the $Z'$ contribution only). 
Due to the stronger dependence on $a$, the $B_s$-mixing constraints are, however, unproblematic for smaller values of $a$, and actually in agreement with the whole $2\,\sigma$ range for $C_9$ for $a\leq 1/3$. Values like $a = 1/2 $ or $a=1/3$ and $m_{Z'}/g' \simeq 2$--$4\TeV$ can therefore easily lead to the required $C_9$ contribution necessary to explain \BKs and $R(K)$ (Fig.~\ref{fig:VEVvsA}).
Note that for these statements we assumed $m_A=m_H$, i.e. only took the $Z'$ contribution to $B_s$--$\overline{B}_s$ mixing into account. However, for $m_A<m_H$ the bounds get weakened, while they become stronger for  $m_A>m_H$ due to the (destructive) constructive interference of the $H$ ($A$) contribution with the $Z'$ and the SM one.

For $a\leq 1/3$ and $m_{Z'}/g' = \mathcal{O}(\TeV)$, direct searches at the LHC cut into the $m_{Z'}$--$g'$ parameter space that is unconstrained by low-energy processes. We then need $m_{Z'} \gtrsim 2.55\TeV$ ($2.46\TeV$) for $a=1/2$ ($1/3$) if we want to explain \BKs and $B\to K\mu^+\mu^-/B\to K e^+ e^-$ within $2\sigma$ (Fig.~\ref{ATLAS_limits})\footnote{Note that the ATLAS constraints can also be evaded for $m_{Z'}\ll \TeV$ with much smaller $g'$ (Fig.~\ref{ATLAS_limits}). For $a=1/2$ ($1/3$), this would require $m_{Z'}<300\GeV$ ($400\GeV$) and $g'< 0.06$ ($0.1$), not necessarily compatible with the approximations used above, so we omit a discussion for now.}.
This also implies a lower limit on the gauge coupling $g'\gtrsim 0.55$ ($0.65$) for $a = 1/2$ ($1/3$), resulting in a $U(1)'$ Landau pole below $10^{15}\GeV$ ($3 \times 10^{12}\GeV$).

We remark that the dominant flavour violation in the $b$--$s$ sector also induces the decay $h\to bs$, with branching ratio of order $10^{-3} \cos^2 (\alpha-\beta) \tan^2\beta$. While generically unobservably small due to the $B_s$-mixing constraints in Eq.~\eqref{eq:DeltaMB_from_scalars}, it can be large if the $A$ contribution to $\Delta m_{B_s}$ takes just the right value.

\section{Extension to three scalar doublets}
\label{sec:3HDM}

Above we considered a 2HDM with a horizontal $U(1)'$ gauge symmetry that leads to flavour-violating couplings of $h$ and $Z'$ to quarks and can successfully explain the anomalies in $B \to K^*\mu^+\mu^-$ and $R(K)$. In this section we will additionally aim at explaining the tantalizing hint for $h\to\mu\tau$ from CMS~\cite{Khachatryan:2015kon} (see Eq.~\eqref{h0taumuExp}), which violates $L_\mu-L_\tau$ by two units. The signal can be accommodated in gauged $U(1)_{L_\mu-L_\tau}$ models by breaking the symmetry with a scalar doublet $\Psi_3$ carrying $|Q'| = 2$~\cite{Heeck:2014qea,Crivellin:2015mga}. Since we cannot set $|a| = 2$ in our 2HDM from above if we want to explain the LHCb anomalies (see Fig.~\ref{fig:VEVvsA}), we have to introduce a third doublet that carries $|Q'| = 2$.
Thus, in total three scalar doublets,
\begin{align}
\Psi_j \equiv \matrixx{\psi_j^+\\ (v_j + \psi_j^{0,{\rm R}} - i \psi_j^{0,{\rm I}})/\sqrt{2}} , \quad j = 1,2,3\,,
\end{align}
with $v \equiv \sqrt{v_1^2+v_2^2+v_3^2} \simeq 246 \GeV$, are introduced to generate viable fermion masses and mixing, as well as the desired lepton-flavor violation in $h\to\mu\tau$. They have the $U(1)'$ charges
\begin{align}
Q'(\Psi_1)=-a\,, &&
Q'(\Psi_2) = 0\,, &&
Q'(\Psi_3) = -2\,,
\end{align}
respectively. We need again (at least) two singlet scalars to break the $U(1)'$ spontaneously above the electroweak scale. The same arguments from above apply regarding the need for more scalars if $a<1/3$ is taken.

For simplicity we will again assume a negligible mixing of the doublets $\Psi$ and singlets $\Phi$, as in Sec.~\ref{sec:2HDM}, which makes sense in the limit $\langle\Phi_j\rangle, m_{\Phi_j} \gg \langle\Psi_i\rangle, m_{\Psi_i} $. In this case, we effectively end up with a restricted 3HDM at low energies with a scalar potential
\begin{align}
\begin{split}
V &\simeq \sum_{j = 1,2,3} \left( m_j^2 |\Psi_j|^2 + \frac{\lambda_j}{2} |\Psi_j|^4 \right)\\
&\quad + \lambda_{12} |\Psi_1|^2 |\Psi_2|^2+ \lambda_{13} |\Psi_1|^2 |\Psi_3|^2+ \lambda_{23} |\Psi_2|^2 |\Psi_3|^2\\
&\quad + \tilde\lambda_{12} |\Psi_1^\dagger \Psi_2|^2+ \tilde\lambda_{13} |\Psi_1^\dagger \Psi_3|^2+ \tilde\lambda_{23} |\Psi_2^\dagger \Psi_3|^2\\
&\quad +\left( m_{12}^2 \Psi_1^\dagger \Psi_2 + m_{23}^2 \Psi_2^\dagger \Psi_3 + \hc\right) .
\end{split}
\end{align}
The mass terms $m_{ij}^2$ of the last line are of particular importance and are generated by the singlet VEVs $\langle \Phi_j\rangle$; note the absence of a term $m_{13}^2 \Psi_1^\dagger \Psi_3$ and quartic terms like $\Psi_2^\dagger \Psi_1\Psi_3^\dagger \Psi_3$. Even though it may look complicated, the above potential (with 16 real parameters) is already much simpler than the general 3HDM (with 54 parameters) due to our imposed (and essentially softly broken) $U(1)'$ symmetry.

Compared to the 2HDM from Sec.~\ref{sec:2HDM}, we now need two angles $\beta$ to parametrize the ratios $\langle\Psi_2\rangle/\langle\Psi_{1,3}\rangle$, and three angles $\alpha$ to describe the mixing of the three CP-even fields $h$, $H_{1,2}$. In addition, also the CP-odd fields $A_{1,2}$ and the charged fields $H^+_{1,2}$ mix among each other. 

Assuming a CP-conserving potential, i.e.~real $m_{12,23}^2$, the minimization conditions give
\begin{align}
m_2^2 \simeq - \frac{\lambda_2 v_2^2}{2} \,, && m_{12}^2 \simeq - \frac{2 m_1^2 v_1}{v_2}\,, && m_{2 3}^2 \simeq - \frac{2 m_3^2 v_3}{v_2} \,,
\end{align}
in the limit $v_2\gg v_{1,3}$. This limit is particularly useful in this 3HDM because it allows for approximations leading to rather simple analytic expressions. It is valid because we already know from the above analysis that $v_1\ll v_2$, and from Refs.~\cite{Heeck:2014qea,Crivellin:2015mga} that $v_3\ll v_2$, i.e.~we are in the ``large $\tan\beta$'' region in both sectors.

For the pseudo-scalars, the would-be Goldstone boson eaten by the $Z$ is given by $G^0 = \sum_j v_j \psi_j^{0,{\rm I}}/v$; the pseudo-scalars orthogonal to $G^0$ further mix due to a complicated $2\times2$ mass matrix. 
We write $(A_1,G^0,A_2) = U_A (\psi_1^{0,{\rm I}},\psi_2^{0,{\rm I}},\psi_3^{0,{\rm I}})$ with the parametrization
\begin{align}
\begin{split}
U_A &= \matrixx{-1 & & \\ & 1 & \\ & & -1}
 \matrixx{1 & & \\ & \cos \eta_{23} & \sin\eta_{23}\\& -\sin\eta_{23} & \cos\eta_{23}}\\
&\times\matrixx{\cos\eta_{13} & & \sin\eta_{13}\\ & 1 & \\ -\sin\eta_{13} & & \cos\eta_{13}}
\matrixx{\cos\eta_{12} & -\sin\eta_{12} & \\ \sin\eta_{12} & \cos\eta_{12} & \\ & & 1 } .
\end{split}
\end{align}
For $v_{1,3}\ll v_2$, the angles are approximately given by
\begin{align}
\eta_{12} &\simeq \frac{v_1}{v}\,, \quad 
\eta_{23} \simeq \frac{v_3}{v}\,, \\
\eta_{13} &\simeq \frac{v_1 v_3}{v^2} \frac{2 m_3^2 + (\lambda_{23}+\tilde\lambda_{23})v^2}{2 (m_1^2-m_3^2) + (\lambda_{12} + \tilde\lambda_{12} - \lambda_{23} - \tilde\lambda_{23})v^2} \,.
\end{align}
Here, $\eta_{13}$ is the most complicated angle, but turns out to be small due to $v_1 v_3\ll v^2$. 
While it is obvious that the $12$ and $23$ mixing is small in the limit $v_{1,3} \ll v_2$, it is less obvious why the $13$ mixing is so small. The reason is the absence of a term $m_{13}^2 \Psi_1^\dagger \Psi_3$ in the potential, which implies that all $13$ mixing originates from the products $v_1 v_3$ of VEVs. We will neglect the mixing in the $13$ sector in the following, which drastically simplifies matters. 
In the limit $v_{1,3}\ll v_2$, we then define the two beta angles $\beta_{12,23} \simeq \pi/2 - \eta_{12,23}$, i.e.~$\tan\beta_{ij}\simeq 1/\eta_{ij}$. $\beta_{12}$ then corresponds to the $\beta$ angle from above (Sec.~\ref{sec:2HDM}), while $\beta_{23}$ corresponds to the $\beta$ angle from Refs.~\cite{Heeck:2014qea,Crivellin:2015mga} and is hence expected to satisfy $\tan\beta_{23}\gg 1$ (to be quantified below).

The charged fields $\psi_j^+$ have the same mixing pattern to linear order in $v_{1,3}$: $( H_1^+, G^+,H_2^+) \simeq U_A (\psi_1^+,\psi_2^+,\psi_3^+)$, with would-be Goldstone boson $G^+$.
Finally, the CP-even fields $(\psi_1^{0,{\rm R}},\psi_2^{0,{\rm R}},\psi_3^{0,{\rm R}})$ have a symmetric mass matrix
\begin{widetext}
\begin{align}
M^2_\mathrm{S} \simeq 
\matrixx{m_1^2 + \tfrac12 (\lambda_{12}+\tilde\lambda_{12})v^2 & \left(\tfrac12 (\lambda_{12}+\tilde\lambda_{12})v - m_1^2/v\right) v_1 & (\lambda_{13}+\tilde\lambda_{13})v_1 v_3 \\ \cdot & \lambda_2 v^2 & \left(\tfrac12 (\lambda_{23}+\tilde\lambda_{23})v - m_3^2/v\right) v_3 \\ \cdot & \cdot & m_3^2 + \tfrac12 (\lambda_{23}+\tilde\lambda_{23})v^2}.
\end{align}
\end{widetext}
The mixing is obviously again suppressed by $v_{1,3}/v$; in particular the $13$ mixing is small. We write  $(H_1,h, H_2) = U_H (\psi_1^{0,{\rm R}},\psi_2^{0,{\rm R}},\psi_3^{0,{\rm R}})$ with 
\begin{align}
U_H \simeq \matrixx{1 & & \\ & \cos \alpha_{23} & -\sin\alpha_{23}\\& \sin\alpha_{23} & \cos\alpha_{23}}
\matrixx{\cos\alpha_{12} & \sin\alpha_{12} & \\ -\sin\alpha_{12} & \cos\alpha_{12} & \\ & & 1 } ,
\end{align}
where we already neglected the small $13$ mixing. The remaining two scalar mixing angles take the form
\begin{align}
\alpha_{12} &\simeq -\frac{v_1}{v} \frac{2 m_1^2- (\lambda_{12}+\tilde\lambda_{12}) v^2}{2 m_1^2 - 2\lambda_2 v^2 + (\lambda_{12}+\tilde\lambda_{12})v^2} \,,\\
\alpha_{23} &\simeq -\frac{v_3}{v} \frac{2 m_3^2- (\lambda_{23}+\tilde\lambda_{23}) v^2}{2 m_3^2 - 2\lambda_2 v^2 + (\lambda_{23}+\tilde\lambda_{23})v^2} \,.
\end{align}

Keeping in mind that the second and third scalar doublets hardly mix in the limit of interest, our 3HDM simplifies significantly, especially taking into account that $\Psi_3$ only couples to leptons and $\Psi_1$ only to quarks. Essentially, our model looks like the 2HDM from Ref.~\cite{Heeck:2014qea} in the lepton sector, \emph{and like a separate 2HDM in the quark sector.} This means we can describe the lepton sector as a type-I-like 2HDM with an angle $\beta_{23}$ and $\alpha_{23}$, and the quark sector as a type-I-like 2HDM with angle $\beta_{12}$ and $\alpha_{12}$. The scalar doublet $\Psi_2$ is mostly SM-like and will inherit off-diagonal couplings to $\mu\tau$ from $\Psi_3$ and to $t u$, $t c$, $d b$, $s b$ from $\Psi_1$.

In this limit, our 3HDM is parametrized by the masses ($m_h$, $m_{H_{1,2}}$, $m_{H_{1,2}^+}$, $m_{A_{1,2}}$), the vacuum-angles $\beta_{12,23}$, and the mixing angles $\alpha_{12,23}$ for the CP-even scalars.

\subsection{Quark masses and couplings}

For the quark masses and couplings, nothing changes compared to the 2HDM above; we simply replace $\tan\beta \to \tan\beta_{12}\simeq v/v_1$ and $\alpha\to \alpha_{12}$, keeping in mind that this only works in the limits of large $\tan\beta_{ij}$ and small $\alpha_{ij}$.
In particular, we can again easily resolve the anomalies in \BKs and $B\to K\mu^+\mu^-/B\to K e^+ e^-$ using the flavour violating $Z'$ couplings as discussed above. As we will see below, the additional constraints from the lepton sector (compared to the 2HDM discussed before) do not interfere with this solution. In fact, the additional resolution of $h\to\mu\tau$ makes possible a prediction for $\tau\to3\mu$, depending on $C_9$.

\subsection{Lepton masses and couplings}

Lepton-flavour violation arises as in Ref.~\cite{Heeck:2014qea}, with $\Psi_3$ playing the role of the non-SM doublet. The Yukawa couplings are given by
\begin{align}
\begin{split}
{\cal L}\; &=  - {\bar L_f}\left( {\xi _{fi}^{\nu_R}{{\tilde \Psi }_3} + Y_{fi}^{\nu_R}{{\tilde \Psi }_2}} \right){\nu_{R,i}} \\
&\quad - {\bar L_f}\left( {\xi _{fi}^e{\Psi_3} + Y_{fi}^e{\Psi _2}} \right){e_i} + \hc
\end{split}
\label{eq:leptonyukawas}
\end{align}
$Y^{\nu_R}$ and $Y^e$ are diagonal due to the $U(1)'$ symmetry, while $\xi^{{\nu_R},e}$ are given by
\begin{align}
\xi^{\nu_R} = \matrixx{0 & 0 & 0 \\ 0 & 0 & \xi_{23}\\0&0&0} , &&
\xi^{e} = \matrixx{0 & 0 & 0 \\ 0 & 0 & 0\\0& \xi_{\tau\mu} &0} .
\end{align}
The right-handed neutrino mass matrix takes again the form of Eq.~\eqref{eq:right-handed_mass} with two vanishing entries. However, due to the non-diagonal Dirac matrices, the active neutrino mass matrix no longer features two vanishing minors but only one, softening the fine-tuning to obtain valid mixing parameters. We again expect a quasi-degenerate neutrino mass spectrum and a close-to-maximal atmospheric mixing angle~\cite{Heeck:2014qea} but lose the very specific predictions we obtained in the 2HDM from Sec.~\ref{sec:neutrinos}.

Diagonalization of the charged-lepton mass matrix requires only a small $23$ rotation of $(\mu_R,\tau_R)$ by an angle $\theta_R \simeq \cos \beta_{23}\xi_{\tau\mu} v/\sqrt{2} m_\tau$, while the left-handed angle is suppressed by $m_\mu/m_\tau$. The LFV coupling of $h$ is then approximately given by
\begin{align}
{\cal L} \ \supset \ -\theta_R \frac{m_\tau}{v} \frac{\cos (\alpha_{23}-\beta_{23})}{\cos\beta_{23}\sin\beta_{23}} \, \overline{\tau} P_R \mu h + \hc
\label{eq:LFVhiggscoupling}
\end{align}
A non-zero $\theta_R$ hence induces the decay $h\to\mu\tau$, while any charged LFV involving electrons is forbidden (in the limit of zero neutrino masses).
The scalars -- $H_2$, $A_2$, and $H_2^+$ -- also couple off diagonally to leptons, but their effect in LFV observables is small~\cite{Heeck:2014qea}.
The $Z'$ couplings are given by
\begin{equation}
{\cal L}\ \supset \ g'\left({{{\bar e}}_i}{\gamma ^\mu }{P_L}{{e}_j}{Z'_\mu }\Gamma_{ij}^{{e}L} + {{{\bar e}}}_i{\gamma ^\mu }{P_R}{{e}_j}{Z'_\mu }\Gamma_{ij}^{{e}R}\right) ,
\end{equation}
with $\Gamma^{e L} \simeq \diag (0,1,-1)$ and
\begin{align}
\Gamma _{fi}^{e R} \simeq
\matrixx{ 0 & 0 & 0\\ 0 & 1 & 2 \theta_R \\ 0 & 2\theta_R & -1} .
\end{align}
Again, a non-zero $\theta_R$ induces LFV in the $\mu$--$\tau$ sector, mediated by the $Z'$.

\subsection{Lepton-flavour violation}

Concerning LFV we can directly rely on the analysis of Refs.~\cite{Heeck:2014qea,Crivellin:2015mga}. The branching ratio for $h\to\mu\tau$ reads
\begin{align}
\begin{split}
{\rm Br}&\left[h\to\mu\tau\right] \simeq \dfrac{m_{h}}{8\pi \Gamma_{\rm SM}} \left| {\Gamma^{h}_{\tau\mu}} \right|^2 \\
&\simeq 1\% \, \left(\frac{\theta_R}{0.1}\right)^2 \left(\frac{\cos(\alpha_{23}-\beta_{23})}{0.2}\right)^2 \left(\frac{\tan\beta_{23}}{20}\right)^2,
\end{split}
\end{align}
where $\Gamma_{\rm SM}\simeq 4.1 \MeV$ is the decay width in the SM for a $125\GeV$ Brout--Englert--Higgs boson and $\Gamma^{h}_{\tau\mu}$ is given by the $\overline{\tau}P_R \mu h$ prefactor in Eq.~\eqref{eq:LFVhiggscoupling}.

LFV mediated by $Z'$ most importantly induces the decay $\tau\to 3\mu$, with $\tau\to \mu \gamma$ suppressed by an additional factor $2\alpha_\mathrm{EM}/\pi$~\cite{Crivellin:2015mga}. The branching ratio is given by
\begin{align}
{\rm{Br}}\left[ {\tau  \to 3\mu } \right] \simeq \frac{{m_\tau ^5 \theta_R^2 }}{{128{\pi ^3}{\Gamma _\tau }}}\frac{{{{g'}^4}}}{{m_{Z'}^4}}\simeq 10^{-8} \left(\frac{\theta_R}{0.1}\right)^2 \left(\frac{6.6\TeV}{m_{Z'}/g'}\right)^4 ,
\end{align}
which has to be compared to the current upper limit of $1.2\times 10^{-8}$ at $90\%$~C.L.~which is obtained from combining data from Belle and BaBar~\cite{Amhis:2014hma}. 
This limit can most likely be improved by an order of magnitude to $10^{-9}$ in the future~\cite{Aushev:2010bq}.

\begin{figure}[t]
\includegraphics[width=0.46\textwidth]{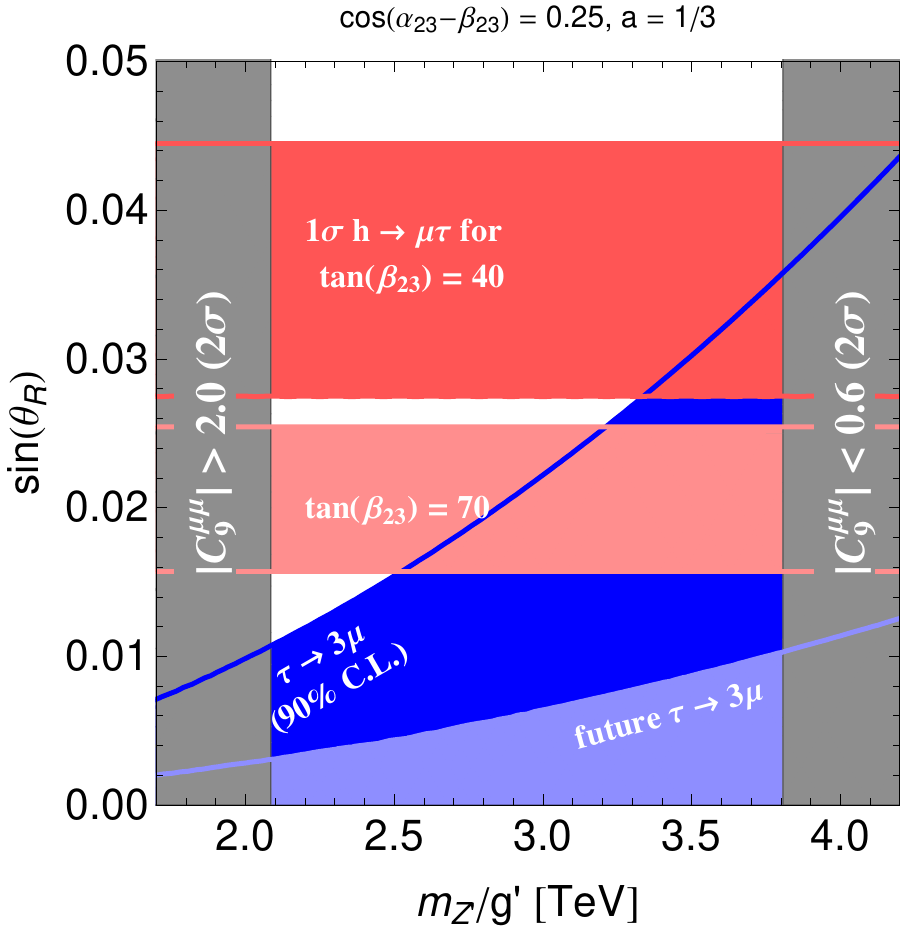}
\caption{Allowed regions in the $m_{Z'}/g'$--$\sin (\theta_R)$ plane for $a = 1/3$: the horizontal stripes correspond to $h\to\mu\tau$ ($1\sigma$) for $\tan\beta_{23}=70,\,40$ and $\cos (\alpha_{23}-\beta_{23})=0.25$, and (light) blue stands for (future) $\tau\to 3\mu$ limits at $90\%$~C.L. The gray regions are excluded by the $2\,\sigma$ range for $C_9^{\mu\mu}$ (see \eq{eq:C9fit}). In this range, ATLAS limits constrain $m_{Z'}\gtrsim 2.5\TeV$ (see Fig.~\ref{ATLAS_limits}).}
\label{fig:vevplot}
\end{figure}

In the previous sections, we have seen that a resolution of the $B$-meson anomalies -- indicated through a non-zero $C_9$ (\eq{eq:C9fit}) -- requires $m_{Z'}/g'$ to be in the TeV range (Fig.~\ref{fig:VEVvsA}).
In Fig.~\ref{fig:vevplot} we show the exclusion limits from $\tau\to 3\mu$ together with the preferred region for $h\to\mu\tau$ and the $C_9$ constraints on $m_{Z'}/g'$. The important part is the \emph{upper} limit on $m_{Z'}/g'$ from $C_9$.
With a non-zero value for $\theta_R$ required by $h\to\mu\tau$, we can then predict a rate for $\tau\to 3\mu$ mediated by the $Z'$.
For this we express $m_{Z'}/g'$ in terms of $C_9$ and $\theta_R$ in $\rm{Br} [h\to\mu\tau]$ to arrive at
\begin{align}
{\rm{Br}}\left[ {\tau  \to 3\mu } \right] \simeq 4.6 \times 10^{-5} \frac{C_9^2 \cos^2\beta_{23} \sin^2\beta_{23}}{a^2 \cos^2 (\alpha_{23}-\beta_{23})} \rm{Br} [h\to\mu\tau] \,.
\end{align}
We remind the reader that the angles $\alpha_{23}$ and $\beta_{23}$ do not correspond to the 2HDM angles from Sec.~\ref{sec:2HDM} but to those from Refs.~\cite{Heeck:2014qea,Crivellin:2015mga}.
Using the $2\sigma$ \emph{lower} limits on $C_9$ (\eq{eq:C9fit}) and $h\to\mu\tau$ (\eq{h0taumuExp}), as well as the LHC constraint $|\cos (\alpha_{23}-\beta_{23})| \leq 0.4$~\cite{Dumont:2014wha,Dumont:2014kna}, we can predict
\begin{align}
{\rm{Br}}\left[ {\tau  \to 3\mu } \right] \gtrsim 9.3\times 10^{-9} \left(\frac{10}{\tan \beta_{23}}\right)^2 ,
\label{eq:prediction_for_tau_decay}
\end{align}
working in the large $\tan\beta_{23}$ limit and setting $a = 1/3$. The current bound is then $\tan\beta_{23}\gtrsim 9$, while the future reach goes above $\tan\beta_{23} \sim 30$. Using the $1\sigma$ limits for $C_9$ and $h\to\mu\tau$ gives a current (future) bound of $30$ (104) on $\tan\beta_{23}$. This is much stronger than the prediction of Ref.~\cite{Crivellin:2015mga} in a model with vector-like quarks, where $1\sigma$ limits only implied a future reach up to $\tan\beta\sim 60$ (using the updated value for $h\to\mu\tau$ from \eq{h0taumuExp}). The 3HDM with gauged horizontal $U(1)'$ charges studied here is hence more tightly constrained than the 2HDM with vector-like quarks~\cite{Crivellin:2015mga}.

Equation~\eqref{eq:prediction_for_tau_decay} is the main prediction of the simultaneous explanation of the $B$-meson anomalies in connection with $h\to\mu\tau$. Note that in addition to the $m_{Z'}/g'$ limits from $C_9$, ATLAS constrains $m_{Z'}$ vs.~$g'$ (Fig.~\ref{ATLAS_limits}). For the parameters in Fig.~\ref{fig:vevplot}, this imposes the additional bound $m_{Z'}\gtrsim 2.5\TeV$ (or $g'\gtrsim 0.65$), which puts the $U(1)'$ Landau pole below roughly $3\times 10^{12}\GeV$ for $a=1/3$.

\section{Conclusions and Outlook}
\label{sec:conclusion}

In this paper we proposed a model with multiple scalar doublets and a horizontal $U(1)'$ gauge symmetry in which all three LHC anomalies in the flavour sector ($B\to K^*\mu^+\mu^-$, $R(K)$ and $h\to\mu\tau$) can be explained simultaneously. Compared to previous explanations, our model does not require vector-like quarks charged under the new gauge group. The spontaneously broken anomaly-free $U(1)'$ gauge symmetry is generated by
\begin{align}
Q' = (L_\mu - L_\tau) - a (B_1 + B_2 - 2 B_3) \,, \quad a\in\mathbb{Q}\,,
\end{align}
which leads to successful fermion-mixing patterns. In particular, it generates a large (small) atmospheric (reactor) mixing angle in the lepton sector and explains the almost decoupled third quark generation. The universal charges the quarks of the first two generations allow for the generation of the Cabibbo angle without dangerously large effects in Kaon mixing, and the neutralness of electrons under the $U(1)'$ symmetry softens constraints without fine-tuning. 

The observed quark mixing of the CKM matrix requires the $U(1)'$ to be broken with a second scalar doublet with $U(1)'$ charge $-a$, which leads to flavour-violating couplings of the $Z'$ and of the scalars, giving simultaneously a natural explanation for the smallness of $V_{ub}$ and $V_{cb}$. Scalar contributions to $B_s$--$\bar B_s$ mixing typically require $\alpha-\beta\simeq \pi/2$, which is, however, relaxed for $m_A<m_H$. The anomalies in $B\to K^*\mu^+\mu^-$ and $R(K)$ can be explained with a TeV-scale $Z'$ boson and $a<1$ while satisfying $B_s$--$\bar B_s$-mixing constraints and limits from direct $Z'$ searches at the LHC. Future LHC and FCC (Future Circular Collider) searches are very interesting for our model as they might strengthen the current limits or lead to the discovery of the $Z'$ boson.

Introducing a third scalar doublet, with $U(1)'$ charge $-2$, gives rise to the decay $h\to\mu\tau$ in complete analogy to Refs.~\cite{Heeck:2014qea,Crivellin:2015mga}. Together with the large $Z'$ effect necessary to resolve $B\to K^*\mu^+\mu^-$ and $R(K)$, the decay $h\to\mu\tau$ then allows us to predict a rate for $\tau\to 3\mu$, depending on $\tan\beta$ and $\cos(\alpha-\beta)$, potentially measurable in future experiments.

\medskip
\acknowledgments{We thank Gennaro Corcella for useful discussions concerning LHC bounds on $Z'$ models and Ulrich Haisch for useful comments on the manuscript. A.~Crivellin is supported by a Marie Curie Intra-European Fellowship of the European Community's 7th Framework Programme under contract number PIEF-GA-2012-326948. G.~D'Ambrosio acknowledges the partial support my MIUR under the project number 2010YJ2NYW. The work of J.~Heeck is funded in part by IISN and by Belgian Science Policy (IAP VII/37).}

\medskip
\emph{Note added}: During the publication process of this article, new LHCb results were presented at the \emph{Rencontres de Moriond Electroweak Session 2015} which hint at a confirmation of the anomaly in $B\to K^* \mu^+\mu^-$. The global fit now prefers new physics in $C_9^{\mu\mu}$ over the Standard Model by $4.3\sigma$~\cite{Altmannshofer:2015sma}.

\medskip
\appendix
\section{Other horizontal symmetries}
\label{sec:appendix}

Demanding a universal $U(1)'$ quark coupling to the first two generations and a good flavor symmetry in the lepton sector does not uniquely single out our model with $B_1+B_2-2 B_3$ and $L_\mu-L_\tau$. It is well known that, besides $L_\mu-L_\tau$ (which is connected to quasi-degenerate neutrinos), the symmetries $L_e$ and $L_e-L_\mu -L_\tau$ are good zeroth-order approximations for a neutrino mass matrix with normal and inverted hierarchy, respectively~\cite{Choubey:2004hn}. Since these two are anomalous, one can consider $B-3 L_e$~\cite{Ma:1998zg, Salvioni:2009jp} or $B+ 3 (L_e-L_\mu-L_\tau)$~\cite{Lee:2010hf, Heeck:2012cd} as well-motivated gauge symmetries. With non-universal quark charges -- but universal in the first two generations -- we can consider $B_3 - L_e$ or $B_3+ (L_e-L_\mu-L_\tau)$ as anomaly-free gauge symmetries that provide a successful neutrino mixing, single out the third quark generation, and lead to LFV in both lepton and quark sectors. Quite analogously one can consider the lepton symmetries from Ref.~\cite{Araki:2012ip} with a non-universal quark charge, which lead to predictive texture zeros and vanishing minors in the neutrino mass matrix; for example, $3 B_3 + L_e - 3 L_\mu - L_\tau$ generates the texture zeros $(m_\nu)_{11} = 0 = (m_\nu)_{13}$ after seesaw. While all these symmetries are interesting in their own right, they are not useful for our purpose because the $Z'$ coupling to quarks compared to muons is rather large (and not adjustable), so it becomes difficult to generate a large $C_9$ without violating $B$--$\bar{B}$-mixing bounds. In addition, any $Z'$ that couples to electrons unavoidably suffers from stringent LEP constraints~\cite{Heeck:2014zfa}. 
The horizontal gauge symmetry $(L_\mu - L_\tau) - a (B_1 + B_2 - 2 B_3)$ chosen in this paper is hence a remarkably good choice to address the existing hints for flavour violation.

\bibliography{BIB}

\end{document}